\documentclass[aps,prb,twocolumn,notitlepage,showpacs,superscriptaddress,10pt]{revtex4-1}%
\usepackage{graphicx}
\usepackage{amsmath}
\usepackage{bm}
\usepackage{amssymb}
\usepackage{color,soul}
\usepackage{amsfonts}%
\usepackage[caption=false]{subfig}
\usepackage{comment}
\usepackage{color,soul}
\usepackage[dvipsnames]{xcolor}
\usepackage{ulem}

\usepackage{float}
\usepackage{tabularx}
\usepackage{array}   
\newcolumntype{L}{>{$}l<{$}} 
\newcolumntype{R}{>{$}r<{$}} 
\newcolumntype{C}{>{$}c<{$}} 
\usepackage{comment}

\newcommand{\RN}[1]{%
  \textup{\uppercase\expandafter{\romannumeral#1}}%
}

\usepackage[utf8]{inputenc}

\usepackage{hyperref}
\usepackage[capitalize]{cleveref}

\begin{document}

\title{Fate of Bosonic Topological Edge Modes in the Presence of Many-Body Interactions}

\author{Niclas Heinsdorf}
\affiliation{Max-Planck-Institut für Festkörperforschung, Heisenbergstrasse 1, D-70569 Stuttgart, Germany}
\affiliation{Department of Physics and Astronomy \& Stewart Blusson Quantum Matter Institute,
University of British Columbia, Vancouver BC, Canada V6T 1Z4}

\author{Darshan G. Joshi}%
\affiliation{Tata Institute of Fundamental Research, Hyderabad 500046, India}

\author{Hosho Katsura}
\affiliation{Department of Physics, Graduate School of Science, The University of Tokyo, 7-3-1 Hongo, Tokyo 113-0033, Japan}
\affiliation{Institute for Physics of Intelligence, The University of Tokyo, 7-3-1 Hongo, Tokyo 113-0033, Japan}
\affiliation{Trans-scale Quantum Science Institute, The University of Tokyo, 7-3-1, Hongo, Tokyo 113-0033, Japan}

\author{Andreas P. Schnyder}%
\affiliation{Max-Planck-Institut für Festkörperforschung, Heisenbergstrasse 1, D-70569 Stuttgart, Germany}

\date{\today}

\begin{abstract}
Many magnetic materials are predicted to exhibit bosonic topological edge modes in their excitation spectra, because of the nontrivial topology of their magnon, triplon, or other quasi-particle band structures. 
However, there is a discrepancy between theory prediction and experimental observation, which suggests some underlying mechanism that intrinsically suppresses the expected experimental signatures, like the thermal Hall current. Many-body interactions that are not accounted for in the non-interacting quasi-particle picture are most often identified as the reason for the absence of the topological edge modes. Here we report  persistent bosonic edge modes at the boundaries of a ladder quantum paramagnet with gapped triplon excitations in the presence of the full many-body interaction. We use tensor network methods to resolve topological edge modes in the time-dependent spin-spin correlations and the dynamical structure factor, which is directly accessible experimentally. We further show that signatures of these edge modes survive even when the non-interacting quasi-particle theory breaks down, discuss the topological phase diagram of the model, demonstrate the fractionalization of its low-lying excitations, and propose potential material candidates.

\end{abstract}

\maketitle

The enormous success of the description of electronic topological phases of matter quickly inspired research that lead to the generalization of the theory to bosonic quasi-particles, like photons~\cite{hosten2008observation, wang2009observation, ben2016photon, haldane2008possible, raghu2008analogs,onoda2004hall},  phonons~\cite{strohm2005phenomenological, kagan2008anomalous,sugii2017thermal, sheng2006theory}, magnons~\cite{katsura_theory_of_thermal_hall,kondo2019three,kondo2021dirac, matsumoto2011rotational, fujimoto2009hall, shindou2013chiral,shindou2013topological,kim2016realization,joshi2018KH} or triplons~\cite{romhanyi2015hall,joshi2017topological,joshi2019z2}. Analogous to the Quantum or Spin Hall effect for electrons, the nontrivial topology of magnetic excitations contributes to unusal responses like the thermal Hall or Spin Nernst effect in the form of bosonic edge currents. 

\begin{figure}
    \centering
    \includegraphics[width=0.7\linewidth]{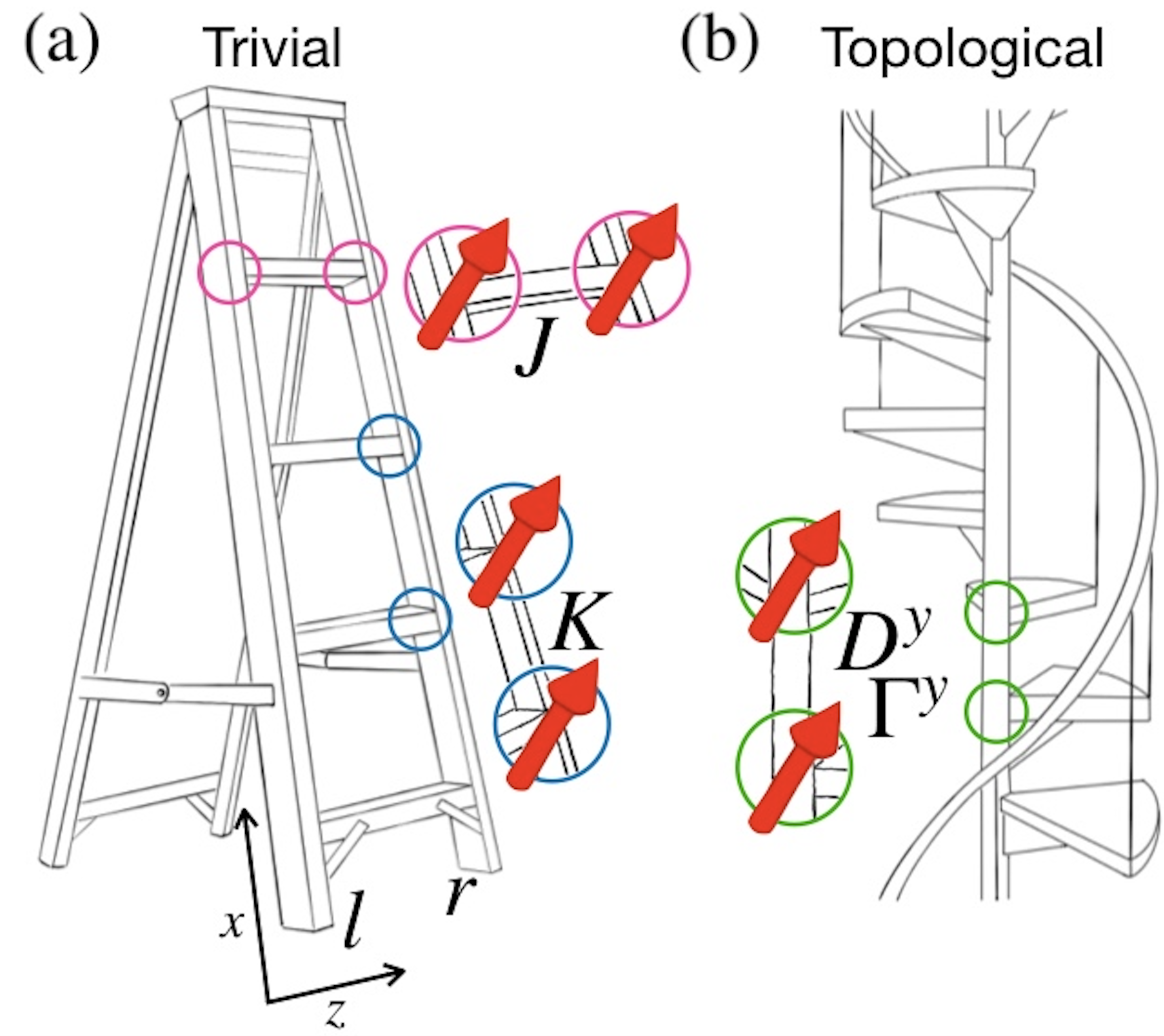}
    \caption{(a) Schematic of the spin ladder. On each rung, two spin-1/2 sites are (strongly) coupled through antiferromagnetic Heisenberg interaction. Along the side rails, the spins interact (weakly) antiferromagnetically. In this limit the spins form singlet pairs along the rungs. (b) The same model but with anti-symmetric (DM) and symmetric (pseudo-dipolar) exchange in the $y$-direction given by $D_y$ and $\Gamma_y$. These terms introduce a winding and open a topological gap in the excitation spectrum.}
    \label{fig:model}
\end{figure}

These topologically protected edge spin waves are a key ingredient for many future spintronic applications, which are a promising alternative to conventional computing, but with a minimal ecological footprint\cite{chumak2015magnon}. There is a multitude of blueprints that make use of their properties to build next-generation devices, like spin-wave diodes, beam splitters, interferometers, and others \cite{li2021topological_applications, matsumoto2011theoretical, wang2018topological, chumak2015magnon}. 

Even though a significant amount of material candidates has been proposed to host nontrivial bosonic excitations~\cite{costa2020topological, li2021topological,katsura_theory_of_thermal_hall, hwang2020topological, laurell2018magnon, matsumoto2011rotational, matsumoto2011theoretical, mook2014magnon, owerre2016topological, han2017spin, romhanyi2015hall, li2016weyl}, their detection remains elusive. While some experiments indeed show signatures of the predicted  edge modes or infer their existence through bulk band topological arguments\cite{zhu2021topological, zhang2021anomalous, mcclarty2017topological, czajka2023planar, hirschberger2015thermal, onose2010observation}, others have not been able to produce evidence of topological quasi-particles~\cite{suetsugu2022intrinsic, cairns2020thermal}, and for most of the proposed candidates a clear indication of their presence is yet to be discovered. 

Many reasons that could make the effects of the topological edge modes fail to materialize have been theorized. Among other things, they include thermal damping, magnon-phonon coupling, and domain formation.
But what is most often conjectured to be the source of the suppression of thermal Hall effect are many-body effects, because they obstruct the application of the standard classification tools which rely on an effectively non-interacting theory. In contrast to their fermionic counterparts, even small many-body interactions~\cite{suetsugu2022intrinsic, chernyshev_review,chernyshev2016damped,habel2023breakdown} or exchange anisotropies that are typically present in realistic models~\cite{thomasen2021fragility} seem to substantially affect bosonic topological transport properties. How interactions affect topological bosonic edge modes in general is yet to be determined, and while many studies go beyond the non-interacting quasi-particle limit\cite{dirac_magnon_honeycomb_ferromagnets_interaction,chernyshev2016damped,winter2017breakdown,sengupta}, only a few studies account for the full many-body interaction\cite{mcclarty_topological_magnons}.


In this work, we report bosonic topological edge modes that are stable in the presence of the full many-body interaction using Density Matrix Renormalization Group (DMRG) and time-evolution methods. We exclude other potential sources of suppression and focus on the consequences of the many-body effects alone. The results presented assume zero temperature and no coupling to any other type of quasi-particles. Moreover, we choose a one-dimensional model, which allows us to study localized edge modes that are not subject to scattering off edge defects present in a real sample.

\textit{Model.}---We consider a quantum $S=1/2$ Heisenberg model on a ladder with spin-orbit interaction and an external magnetic field, as shown schematically in \mbox{Fig.\ \ref{fig:model}}. This model, first considered in Ref. \onlinecite{joshi2017topological}, is given by the following Hamiltonian:

\begin{align}\label{eq:hamiltonian}
    \hat{H} &= \hat{H}_{\mathrm{rung}} + \hat{H}_{\mathrm{rail}} + \hat{H}_{\mathrm{SOC}} + \hat{H}_{Z} , \\
    \hat{H}_{\mathrm{rung}} &= J\sum_i \hat{\bm{S}}_{li} \cdot \hat{\bm{S}}_{ri} ,\\
    \hat{H}_{\mathrm{rail}} &= K\sum_i \left( \hat{\bm{S}}_{li} \cdot \hat{\bm{S}}_{li+1} +  \hat{\bm{S}}_{ri} \cdot \hat{\bm{S}}_{ri+1} \right) ,
    \\
    \hat{H}_{\mathrm{SOC}} &= D_y\sum_{\alpha = l,r} \sum_i \left( \hat{S}^z_{\alpha i}  \hat{S}^x_{\alpha i+1} -  \hat{S}^x_{\alpha i}  \hat{S}^z_{\alpha i+1} \right)  , \nonumber\\
    &+ \Gamma_y \sum_{\alpha = l,r} \sum_i \left( \hat{S}^z_{\alpha i}  \hat{S}^x_{\alpha i+1} +  \hat{S}^x_{\alpha i}  \hat{S}^z_{\alpha i+1} \right) , \\
    \hat{H}_{Z} &= h_y \sum_i \left( \hat{S}^y_{l i} + \hat{S}^y_{r i} \right) ,
\end{align}
where $\hat{\bm{S}}_{\alpha i} = \left(\hat{S}_{\alpha i}^x, \hat{S}_{\alpha i}^y, \hat{S}_{\alpha i}^z \right)^\intercal$ are the usual spin operators with $\alpha = l, r$ corresponding to the left and right rail of the ladder and $i$ running over all rungs at positions $r_i$.

In the limit of strong anti-ferromagnetic rung coupling, this model is the prototypical example of a gapped quantum paramagnet~\cite{schmidiger2012spectral, dagotto1996surprises}. In that limit the ground state of the model is close to a product state of spin singlets 
\begin{align}\label{eq:ground_state}
|\psi_0\rangle \sim \bigotimes_{i=1}^{L}(|\uparrow \downarrow \rangle - |\downarrow \uparrow \rangle)/\sqrt{2},
\end{align}
with low-lying triplet excitations, the corresponding quasiparticles being triplons. In the presence of SU(2) symmetry ($D_y=\Gamma_y=h_y=0$), the triplons are degenerate, but can be split by either applying an external magnetic field or through spin-orbit interactions. 


\begin{figure}
    \centering
    \includegraphics[width=0.9\linewidth]{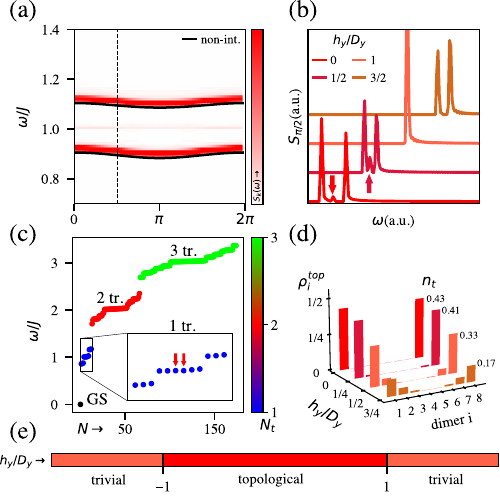}
    \caption{(a) DSF of the topological quantum paramagnet with $L=32$ rungs and $K/J = 0.01$, $D_y/J = \Gamma_y/J = 0.1$ and $h_y = 0$ (corresponding to the red plot in (b)) The solid lines show the spectrum of the effective low-energy models from Ref.~\onlinecite{joshi2017topological} for the same parameters. There is a localized boundary mode at $\omega/J=1$. (b) DSF for different values of $h_y / D_y$ at $k=\pi/2$ (dashed line in (a)). The first two curves lie in the topological phase region (see (e)) with the in-gap modes marked by arrows. At $h_y / D_y = 1$ the gap is closed. The last curve lies in the trivial region and is gapped, with no mode in-between. All peaks are centered at $\omega\approx J$ (and shifted for clarity). (c) Many-body spectrum of a 4-rung system with the same parameters as in (a). The spectrum is separated into sectors of different triplon particle numbers. The color indicates $N_t$. The inset shows the 1-triplon sector with the edge modes marked by arrows. (d) $\rho_i^{\text{top}}$ for different values of $h_y / D_y$, but with $L=8$. The in-gap modes are localized at the two boundaries of the system with fractional particle numbers $n_t$ at each termination. For finite magnetic fields the in-gap mode spreads into the bulk and vanishes at the topological phase transition at $h_y/D_y=1$. (e) The topological phase diagram of the model.}
    \label{fig:new_figure2}
\end{figure}


In the paramagnetic phase of the model, an effective low-energy spectrum can be calculated by first rewriting the model in terms of triplons through bond-operator theory\cite{sachdev1990bond}, and then discarding all resulting terms that are not bilinear, which has been shown to be a controlled approximation \cite{larged1,larged2} for low triplon densities. Within the harmonic approximation, it was shown in Ref.\ \onlinecite{joshi2017topological} that the so-obtained excitation spectrum can be characterized by a topological winding number that classifies the topology of the magnet's bulk and enforces -- if nontrivial -- triplon modes at the ends of the ladder with fractional particle number. The model follows the design philosophy of many topological toy models. That is, SOC is used to open a topological gap by entangling an internal degree of freedom\cite{haldane1988model,kane-mele,fu-kane,bhz,kondo2019z,kondo2019three}. 

The model is suitable to describe the magnetism of potentially many materials and belongs to the well studied class of two-leg ladder compounds. Examples of spin ladder materials are BiCu$_2$PO$_6$\cite{mentre2006structural}, NaV$_2$O$_5$\cite{mila1996exchange,isobe1996magnetic} or multiple cuprate ladders\cite{hiroi1995absence,kohama2012anisotropic,kiryukhin2001magnetic,azuma1994observation}. The recipe for finding topological edge modes in these systems requires mainly two ingredients: (i) strong antiferromagnetic rung coupling (stronger than along the rails), and (ii) spin-orbit interaction, the relative strengths of which can be tuned by doping, gate voltages or structural manipulations like skewing and stretching\cite{wang2016tunable, cheung2016spin, winterfeld2013strain, liu2021semimetallic, esaki2023electric}. The abundance of experimental handles suggests that a topological phase transition might easily be switchable, which again expands the potential applications of these systems in future devices.

\textit{Tensor Networks and Dynamical Response.}---To relate our model directly to an experimentally measurable quantity, we compute the dynamic structure factor (DSF), which is most directly accessed through inelastic neutron scattering\cite{Sturm+1993+233+242, van1954correlations} 
or inverse spin Hall noise spectroscopy (ISHNS)\cite{joshi2018detecting}. Because we are investigating the effect of interactions, and since our model is (quasi) one-dimensional, we use the DRMG algorithm to obtain the exact many-body ground state of the finite spin ladder\cite{white1992density}. 

In contrast to electronic systems, the band topology of bosonic quasiparticles is not a ground state property, so in order to access the low-lying excitations, we apply time-evolution methods\cite{vidal2004efficient, daley2004time, white2004real, haegeman2011time, haegeman2016unifying, zaletel2015time} to compute the odd-part\cite{footnote_oddpart} of the time-dependent spin-spin correlation
\begin{align}
    C^{\gamma \gamma'}_{ij}(t) = \langle \psi_0|\hat{\tilde{S}}^\gamma_i \hat{U}(t)\hat{\tilde{S}}^{\gamma'}_j|\psi_0\rangle,
\end{align}
with $\hat{\tilde{S}}^{\gamma}_i = \hat{S}^{\gamma}_{li} - \hat{S}^{\gamma}_{ri}$ and the unitary time-evolution operator $\hat{U}(t)$. Time-resolved data is typically unavailable for larger (interacting) systems, and the evolution of topological edge states remains largely unexplored. In the appendix on Hilbert space fragmentation, we show that the topological boundary modes have strongly enhanced time coherence and that the model's Hilbert space is fragmented by additional conserved charges in some limit.

The DSF, which encodes the dynamical susceptibility of the paramagnet is defined as the Fourier transform of the spin-spin correlations in space and time
\begin{align}\label{eq:DSF}
    S^{\gamma \gamma'}_k(\omega) = \frac{1}{2\pi L}\int_{-\infty}^\infty dt \ e^{i(\omega - \omega_0) t}\sum_{i,j} e^{i(r_j-r_i)k}C^{\gamma \gamma'}_{ij}(t).
\end{align}
It is a positive and real quantity and can be calculated using DMRG and time-evolution methods\cite{gohlke2018dynamical}. Usually, the state is evolved only for as long as the excitation does not reach the boundary of the system to avoid finite size effects. In addition, by imposing translational invariance, one of the spatial degrees of freedom of $C^{\gamma \gamma'}_{ij}(t)$ can be fixed \cite{sherman2023spectral,bouillot2011statics,laurell2022magnetic,gohlke2018dynamical}. Because we are explicitly studying the excitations in the presence of a boundary, we require the ``full" real-space correlation function, as well as a ``long-enough" time-evolution. $S^{\gamma \gamma'}_k(\omega)$ obeys sum rules\cite{laurell2022magnetic} that we track to diagnose the severity of finite-time and -size effects, which lead to numerical artifacts \mbox{(see supplementary material\cite{supp})}.

We study the $z$-component of the field-susceptible part of \mbox{Eq.\ (\ref{eq:DSF})} which is given by $S_k(\omega) =$\sout{ $S^{xx}_k(\omega) +$}$ S^{zz}_k(\omega)$ to which we simply refer to as DSF from now.

\textit{Results.}---Before moving on to the case of strong SOC, we review the key properties of the model's excitation spectrum in the weakly interacting limit. We set $D_y=\Gamma_y$, and express the Hamiltonian through operators $t^{\gamma}_{i}$ that create a triplon by acting on the singlet state on the $i$th dimer\cite{sachdev1990bond}. The full Hamiltonian is given in the appendix on the model's topological properties, but in particular
\begin{align}\label{eq:triplon_SOC}
    \hat{H}_{\mathrm{SOC}} = D_y\sum_i t^{z\dagger}_iP_iP_{i+1}t^{x}_{i+1} - t^{z\dagger}_iP_it^{x\dagger}_{i+1}P_{i+1} + h.c.,
\end{align}
with $P_i = 1 - \sum_\gamma t_i^{\gamma\dagger}t^\gamma_i$. In Fig.\ \ref{fig:new_figure2} (a) we plot the DSF of the topological paramagnet along with the triplon bands obtained within the harmonic approximation in  Ref.\ \onlinecite{joshi2017topological}. For small SOC the harmonic approximation is accurate, because the density of triplons is small leading to only minor renormalizations from triplon-triplon interactions. In the topologically nontrivial case (small values of magnetic field) an additional mode appears between the two bulk bands that we hereafter refer to as {\it{in-gap mode}}. It is absent in the trivial phase (large values of magnetic field). Fig.\ \ref{fig:new_figure2} (b) shows the topological phase transition at $k=\pi/2$ with the external magnetic field $h_y$ as a tuning parameter. At $h_y = |D_y|$ the gap closes, and further increasing the magnetic field it reopens, but with no topological in-gap mode. We did finite size scaling with ladders of up to $L=64$ rungs to make sure that the in-gap mode retains finite spectral weight. 

As expected, the harmonic approximation and the tensor networks agree on the dispersion relation of the single-triplon bulk excitations. That both predict a topological edge mode is less conspicuous, even for small SOC. While signatures of chiral edge modes (on a cylinder) have been observed using tensor networks\cite{mcclarty_topological_magnons}, the harmonic approximation is not obviously adiabatic. For example, in one-dimensional electron systems finite interactions force a Mott transition through which the topological properties of the system cannot be tracked\cite{iraola2021towards}. In the weakly interacting case, exact diagonalization (ED) on a 4-rung ladder shows that the spectrum has well-separated number of particle sectors. Even though $\hat{H}_{\mathrm{SOC}}$ does not conserve the triplon number $N_t= -\frac{3}{4}JL + J\sum_\gamma \langle t_i^{\gamma\dagger}t^\gamma_i\rangle$, the expectation values are close to integer for this choice of parameters. Because $\hat{H}_{\mathrm{SOC}}$ contains only even powers of $t^{\gamma}_{i}$, single-triplon states -- such as the in-gap mode $|\psi_{\mathrm{top}}\rangle$ -- scatter mainly off states from the three-triplon sector $|\psi_{N}^3\rangle$. This is confirmed by evaluating the matrix elements $\langle \psi_N^3|\hat{H}_{\mathrm{SOC}}|\psi_{\mathrm{top}}\rangle$ which are in general finite, because there is no (hidden) symmetry that prevents the states from interacting.

\begin{figure*}
    \centering
    \includegraphics[width=0.95\linewidth]{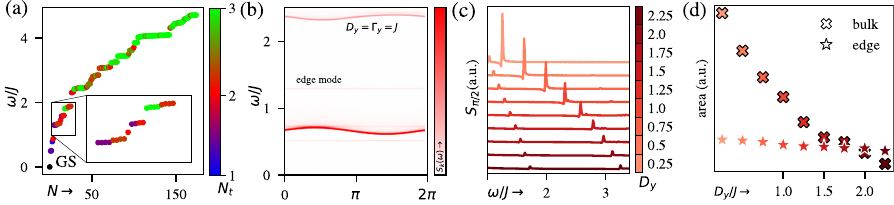}
    \caption{(a) Many-body excitation spectrum with $K = 0.01J$, $h_y = 0.05J$ and strong SOC $D_y = \Gamma_y = J$ on a ladder with $L=4$ rungs. The color indicates the triplon number $N_t$, which is not conserved as indicated by the non-integer values (purple and brown) of many eigenstates, also within the energy range close to the in-gap mode as shown in the inset. (b) DSF for the same parameters with $L=32$. The upper quasi-particle peak has moved up in energy and lost intensity. The edge mode remains flat and is marked by text. (c) DSF at the gap closing point $k=\pi/2$ as a function of energy for different values of $D_y=\Gamma_y$. As a function of SOC, the in-gap mode (left peak) and the bulk band (right peak) move up in energy with a growing separation in energy. (d) Area of the bulk (cross) and the in-gap (star) mode for increasing $D_y=\Gamma_y$. The intensity of the bulk band decreases as it moves up in energy, while the intensity of the in-gap mode is roughly constant.}
    \label{fig:new_figure4}
\end{figure*}

To confirm that the in-gap mode is really localized at the boundaries of the system, we first calculate the local density of states of the in-gap mode $\rho_i^{\text{top}}$ for $L=8$, and then compute the associated particle number by integrating it for the system's left termination (see supplementary material\cite{supp})
\begin{align}\label{eq:nt_integral_formula}
    n_t = \int_1^{L/2} dr \rho_i^{\text{top}}. 
\end{align}
We find two localized peaks at the $x$-boundaries of the ladder with fractional particle number per termination. Our numerical result $n_t \sim 0.43$ deviates slightly from the predicted value of $0.5$~\cite{joshi2017topological} due to spectral leakage caused by the aforementioned finite-size and time effects. For larger values of field, $n_t$ becomes smaller and vanishes at the phase transition at $h_y/D_y = 1$. Details are found in the supplemental material\cite{supp}.

The noninteracting triplon winding number is a topological band property\cite{joshi2017topological}, so the edge modes are naively expected to be stable only as long as there is a notion of ``bands", or in other words as long as triplons provide a suitable quasi-particle description of the ladder's excitation spectrum. We consider three limits of the model's parameter space: (i) large magnetic field $h_y$, (ii) large rail coupling $K$ and (iii) large spin-orbit interaction $D_y$ and $\Gamma_y$. In case (i), the field-polarized phase, the lower triplon modes condenses at $h_y \sim J$ and the system becomes ferromagnetic with magnons as its low-lying excitations. The band structures derived from linear spin wave theory are given in the supplemental material\cite{supp} and are topologically trivial. Case (ii) is the limit of two weakly coupled anti-ferromagnetic Heisenberg chains. The system does not order, even for strong rail coupling ($K \sim 2J$). The antiferromagnetic order parameter is finite also for small values of $K$, but approaches zero as the system size $L$ is increased, as confirmed by finite size scaling~\cite{supp}. 
Case (iii) is the most relevant benchmark of the edge modes' stability that we can provide with the available handles in the model, because here interactions become important (see Eq.\ \ref{eq:triplon_SOC}). In this limit the eigenstates cannot straightforwardly be assigned to a particle number sector, and ED yields non-integer values of $N_t$ as shown in Fig.\ \ref{fig:new_figure4}(a). States with a triplon number close to $N_t=3$ (which we have previously shown to mix with single-triplon states through $\hat{H}_{\mathrm{SOC}}$) now occupy the same energy range as the edge mode, the quasi-particle peak of which remains discernable in the DSF and is marked by text in Fig.\ \ref{fig:new_figure4}(b). The in-gap mode remains flat (localized), and moreover is of the same quasi-particle type as the multi-triplon continuum, which rules out avoided quasi-particle decay that has been shown to enhance quasi-particle lifetimes in some strongly interacting systems\cite{Verresen2019,alex_interaction_stabilized}. Clearly, the harmonic approximation is no longer valid. 
In fact, we find that the harmonic Bogoliubov-de-Gennes Hamiltonian ceases to be positive definite already at just below $D_y=\Gamma_y=J/2$. The intensity of what previously was the upper single-triplon mode is strongly reduced, and we track its evolution together with that of the topological edge mode as a function of SOC at $k=\pi/2$ in Fig.\ \ref{fig:new_figure4}(c). 
The upper peak moves up in energy faster than the in-gap mode, and in contrast to case (i), increasing 
$D_y$ and $\Gamma_y$ does not lead to condensation of the lower mode. We did not find any evidence for a phase transition up to values of $D_y = \Gamma_y = 2.6J$. Analogous to the quasi-particle residue in Fermi liquid theory, which is typically interpreted to be a measure of quasi-particle coherence\cite{tremblay2008refresher,fateflatiron}, we plot the areas under the two peaks shown in Fig.~\ref{fig:new_figure4}(c) as a function of $D_y/J$ in Fig.\ \ref{fig:new_figure4}(d). We find that the topological and bulk modes show qualitatively different behaviors: The bulk-triplon peak falls off quickly, whereas the area under the in-gap mode is roughly constant (up to some spectral leakage, which becomes more severe as $D_y$ is increased) with a crossover at $D_y/J\approx 2$.

\textit{Discussion}---We have demonstrated that, contrary to what the absence of experimental evidence for many predicted materials might suggest, topological bosonic edge modes can persist in the presence of the full many-body interaction. Even though correlations have been identified as the reason for the absence of topological responses in some cases, our work clearly shows that they do not generically suppress bosonic edge modes, and it prompts the question of what other mechanism is responsible. A natural extension of this work is to apply our method to two-dimensional systems wrapped onto a cylinder. Multi-rail ladders\cite{samajdar2019enhanced, kawano2019thermal}, a ferromagnet on a Lieb lattice\cite{cao2015magnon} or the Shastry-Sutherland model\cite{shastry1981exact} are obvious candidates, but also topological magnon systems that are potentially more fragile due to their larger decay phase space. In two dimensions, the edge modes transport the intrinsic part of the thermal Hall response, implying that suppressing edge modes results in an attenuated response. 

The absence of signatures of topological edge modes even in systems with good agreement between theory and experiment for bulk properties suggests that the responsible decay channel might be a surface effect, i.e., it lies at the boundary. In a quasi two-dimensional model the edge modes are not completely localized, but propagate along the boundary of the system. This setup would further allow the implementation of a defect at the boundary and to investigate the stability of the topological bosons in its presence. 

Thermal effects might be another candidate to explain the missing Hall response in Refs.~\onlinecite{suetsugu2022intrinsic,cairns2020thermal}, and indeed strong thermal damping of triplon modes has been observed in a Shastry-Sutherland compound\cite{strong_triplon_dampening}. 
However, the experiments that report an absence of thermal Hall current still find sharp single-triplon quasi-particle peaks, which suggests that it is unlikely that the broadening due to damping would suppress the signal almost entirely. Thermally activated renormalization effects, as reported in CrBr$_3$\cite{thermal_evolution}, cannot be ruled out by our study. However, the zero-temperature DSF does not suggest enhanced renormalization at the gap closing point (for $|h_y| = |D_y|$) compared to other values of $k$, and a temperature-induced topological phase transition, as predicted in CrI$_3$, seems unlikely for low temperatures and no external magnetic field\cite{temperature_induced_phase_transition}.

The typical ``workflow" for finding topological edge modes usually consists of classifying the topology of a translationally invariant model, and then inferring their existence by bulk-boundary correspondence. Even if the symmetry that protects the band topology is broken weakly, the edge states are still expected to be present, as long as the perturbation is smaller than the gap they live in (these approximate symmetries are sometimes called quasi-symmetries\cite{guo2022quasi, shang2023irf}). The chiral symmetry that allows for the definition of a winding number in the noninteracting case is broken by the rail coupling $K$~\cite{joshi2017topological}, and is also clearly broken in the limit of strong SOC. It is a central question of the field how quantum geometric quantities can be defined in the presence of strong interactions, e.g. when bands are substantially broadened\cite{chernyshev2016damped}. Our work circumvents this ambiguity and approaches the problem from ``the other end of the bulk-boundary correspondence" by including the boundary explicitly. Nonetheless, the extension of topological (bulk) classifications to interacting systems is a standing problem, and in recent years considerable effort has been made to find alternative invariants~\cite{gurarie2011single,wang2013topological,wang2012simplified,iraola2021towards,kapustin2020higher,shiozaki2023higher}. These schemes account for the many-body nature of the problem using either a matrix product state or Green's function representation, but are not applicable for the classification of bosonic excitation spectra. Detection of phase transitions using entanglement have been brought forward, but are - even though directly related to the dynamical response of a spin system - not sensitive to topological phase transitions\cite{hauke2016measuring}. It is essential to extend existing or find new methods that are able to capture the topological properties of e.g. dynamical spin responses, and we hope our work inspires fruitful research in that direction.

\nocite{hauschild2018efficient,harris2020array}

\begin{acknowledgments}
We thank A. Nocera and S. M. Winter for helpful discussions. NH acknowledges financial support from the Max Planck Institute for Solid State Research in Stuttgart, Germany and the DAAD JSPS summer program 2022. DGJ acknowledges support from the Department of Atomic Energy, Government of India, under Project Identification No. RTI 4007.
A.P.S. and N.H. acknowledge support by the Deutsche Forschungsgemeinschaft (DFG, German Research Foundation) – TRR 360 – 492547816. H.K. was supported by JSPS KAKENHI Grant No. JP23K25790 and MEXT KAKENHI Grant-in-Aid for Transformative Research Areas A “Extreme Universe” (KAKENHI Grant No. JP21H05191).
\end{acknowledgments}

\setcounter{figure}{0}
\makeatletter
\renewcommand{\fnum@figure}{\figurename~A\thefigure}
\makeatother

\textit{Appendix on bond operator theory}.---
We give an expression for the Hamiltonian in terms of triplon operators, following the supplemental material of Ref.~\onlinecite{joshi2017topological}. 

\begin{figure}
    \centering
    \includegraphics[width=0.9\linewidth]{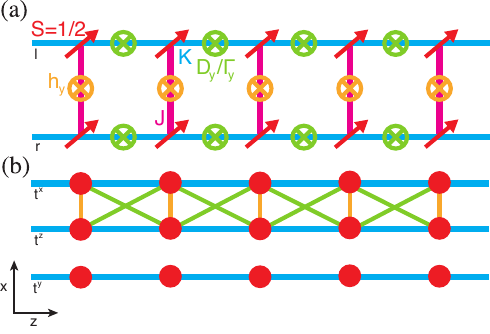}
    \caption{(a) Schematic of the model of the quantum paramagnet from Eq.\ \eqref{eq:hamiltonian}. (b) The same model in the basis of triplon operators $t^\gamma$ up to bilinear terms. $D_y$, $\Gamma_y$ (green) and $h_y$ (orange) couple the $t^x$ and $t^z$ chains along the correspondingly colored bonds. $J$ is a chemical potential. The full Hamiltonian is given in Eq.\ \eqref{aeq:triplon_hamiltonian}.}
    \label{fig:triplon_schematic}
\end{figure}

As in the main text, we consider strong anti-ferromagnetic rung coupling and denote the singlet state as $|t^0\rangle = (|\uparrow \downarrow \rangle - |\downarrow \uparrow \rangle)/\sqrt{2}$ (see Eq.\ \eqref{eq:ground_state}). Then, the triplon operators are defined as 
\begin{align}
    t^{x\dagger}|t^0\rangle &= -(|\uparrow \uparrow \rangle - |\downarrow \downarrow \rangle)/\sqrt{2}, \\
    t^{y\dagger}|t^0\rangle &= i(|\uparrow \uparrow \rangle + |\downarrow \downarrow \rangle)/\sqrt{2}, \\
    t^{z\dagger}|t^0\rangle &= (|\uparrow \downarrow \rangle + |\downarrow \uparrow \rangle)/\sqrt{2},
\end{align}
and the spin operators can be written as 
\begin{align}
    \hat{S}^\gamma_{\alpha i} = \frac{1}{2}\left(\pm it^{\gamma\dagger}_i P_i \mp iP_it^{\gamma}_i - i\epsilon_{\gamma\gamma'\tilde{\gamma}}t^{\gamma'\dagger}_it^{\tilde{\gamma}}_{i} \right),
\end{align}
with $\alpha=l,r$ and $P_i = 1 - \sum_\gamma t_i^{\gamma\dagger}t^\gamma_i$. Then
\begin{align}
    \hat{\bm{S}}_{li} \cdot \hat{\bm{S}}_{ri} = -\frac{3}{4}+ \sum_\gamma t^{\gamma\dagger}_it^{\gamma}_{i}, 
\end{align}
and inserting this into the Hamiltonian in the main text (see Eq.\ \eqref{eq:hamiltonian}) yields
\begin{align}\label{aeq:triplon_hamiltonian}
    \hat{H} = J\sum_\gamma\sum_i &t^{\gamma\dagger}_it^{\gamma}_{i} 
    -\frac{3}{4}JL 
    +ih_y \sum_i \left[ t^{x\dagger}_it^{z}_{i} - t^{z\dagger}_it^{x}_{i} \right]
    \nonumber\\ 
    +\frac{K}{2}\sum_i\sum_{\gamma}  & \left[ t^{\gamma\dagger}_iP_iP_{i+1}t^{\gamma}_{i+1} - t^{\gamma\dagger}_iP_it^{\gamma\dagger}_{i+1}P_{i+1} + h.c. \right]
    \nonumber \\ +\frac{D_y}{2}\sum_i 
    & \Big[ t^{z\dagger}_iP_iP_{i+1}t^{x}_{i+1} - t^{z\dagger}_iP_it^{x\dagger}_{i+1}P_{i+1} \nonumber \\
    -&t^{x\dagger}_iP_iP_{i+1}t^{y}_{i+1} + t^{x\dagger}_iP_it^{z\dagger}_{i+1}P_{i+1} + h.c. 
    \Big]
    \nonumber \\
    +\frac{\Gamma_y}{2}\sum_i 
    & \Big[ t^{z\dagger}_iP_iP_{i+1}t^{x}_{i+1} - t^{z\dagger}_iP_it^{x\dagger}_{i+1}P_{i+1} \nonumber \\
    +&t^{x\dagger}_iP_iP_{i+1}t^{y}_{i+1} - t^{x\dagger}_iP_it^{z\dagger}_{i+1}P_{i+1} + h.c.
    \Big] .
\end{align}
In Fig.\ A\ref{fig:triplon_schematic} we give a schematic of the (bilinear) terms in Eq.\ \eqref{aeq:triplon_hamiltonian}. Because SOC points along the $y$-direction, it only couples the $x$- and $z$-triplon modes. As described in the main text, the triplon number $N_t$ is not conserved. If $D_y=\Gamma_y$, the terms with opposite signs cancel and one obtains an expression for $\hat{H}_{\mathrm{SOC}}$ as given in Eq.\ \eqref{eq:triplon_SOC} in the main text. 

In contrast to the case of topological magnons of ordered magnets~\cite{hirschberger2015thermal,bao2018discovery,elliot2021order,yao2018topological,yuan2020dirac,zhang2020magnonic}, triplon excitations are gapped. As a consequence of this bulk gap the triplons, including the topological edge modes, do not have any low-energy states to decay into. Therefore their lifetime is not reduced significantly even in the presence of weak interactions. By applying an external magnetic field or through spin-anisotropy exchanges, magnon systems can become gapless too. However, a bulk gap induced by these effects is typically much smaller than the rung coupling $J$, which is the relevant spin exchange energy scale that determines the size of the gap in the ladder system.

\textit{Appendix on Hilbert space fragmentation}.--- 
\begin{figure}
    \centering
    \includegraphics[width=0.95\linewidth]{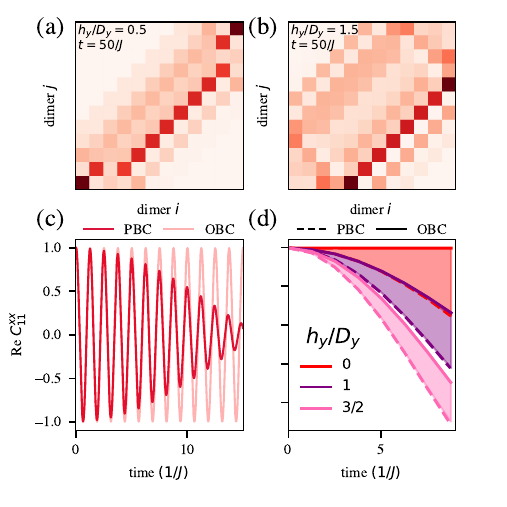}
    \caption{Real part of the spin-spin correlation function at time $t=50/J$ in the (a) trivial and (b) the topological case with $K=0.01J$ and field strengths $h_y/D_y = 0.5$ and $h_y/D_y = 1.5$, respectively, on a ladder with $L=12$ rungs. 
    The topological auto-correlation at this time step has peaks at the boundaries, which are absent in the trivial case. 
    (c)~Time dependence of the boundary auto-correlation $\mathrm{Re}\ C^{zz}_{11}$ for $K=0.01J$, $D_y=\Gamma_y = 0.1J$ and no magnetic field on a ladder with $L=8$ rungs for different boundary conditions. For PBC (no in-gap mode), it decays rapidly whereas for OBC (in-gap mode) it does not decay. (d) The decay envelopes of (c) for different values of magnetic field. For finite field strengths the correlator decays also for OBC, but more slowly than for PBC.   }\label{fig:real_time_real_space_correlations}
\end{figure}
Systems that are useful for storing and processing quantum information are highly sought-after. Typically, the quantum information of a many-body state is destroyed rapidly. Overcoming fast decoherence of quantum states is one of the main challenges of quantum computing, and has driven extensive theoretical interest in finding ways to increase their coherence times using many-body localization~\cite{gornyi2005interacting,nandkishore2015many,altman2015universal,alet2018many,abanin2019colloquium}, prethermalization~\cite{jermyn2014stability, kemp2017long, else2017prethermal}, quantum many-body scarring~\cite{serbyn2021quantum,chandran2023quantum, moudgalya2022quantum} or the states' topological properties\cite{parker2019topologically,amelio2020theory,zapletal2020long,bahri2015localization}. 
Investigating time evolution of topological states - specifically in the presence of defects or nonlinearities\cite{chaunsali2021stability,smirnova2020nonlinear,leykam2016edge} - as well as imaging and detection techniques\cite{goldman2012detecting,goldman2013direct, yao2017theory} are thriving fields of research and crucial for bringing real-world applications on their way.In the following, we show that the boundary mode in the topological quantum paramagnet has enhanced time coherence, which we then argue to be  linked to Hilbert space fragmentation. 

In Figs.\ A\ref{fig:real_time_real_space_correlations}(a) and (b) the real part of $C^{\gamma \gamma'}_{ij}(t)$ for the topological and trivial case at a fixed time step are plotted. The former shows strong correlations that are pinned to the edges of the system, whereas in the latter they are absent. 
In Fig.\ A\ref{fig:real_time_real_space_correlations}(c), we plot the time dependence of $\mathrm{Re}\ C^{xx}_{11}$ at zero magnetic field for periodic (PBC) and open boundary conditions (OBC). For PBC there is no topologically protected edge mode at $i=j=1$, and the excitation decays over time rapidly. In the open system, the edge mode has strongly enhanced time coherence, and we cannot resolve any decay during the time it takes the excitation to hit the boundary at the other end of the ladder. 
Fig.\ A\ref{fig:real_time_real_space_correlations}(d) shows the decay envelope of the excitation for increasing magnetic field for periodic (dashed line) and open (solid line) boundaries. For finite magnetic fields, the boundary correlation starts to decay, and for stronger fields the difference in lifetime for PBC and OBC - as indicated by the shaded area - becomes less pronounced.

There is no abrupt change in behavior at the phase transition $D_y=\Gamma_y=h_y$, which makes it unlikely that the topological properties of the system alone are responsible for the observed behavior. Rather, it seems to imply that the $h_y/D_y=0$ is a special point in the phase diagram. Consider the case of $K=h_y=0$ and $D_y=\Gamma_y$. Then, Eq.\ \eqref{eq:hamiltonian} simplifies to $\hat{H} = \hat{H}_{\mathrm{rung}} + \hat{H}_{\mathrm{SOC}}$ and $\hat{H}_{\mathrm{SOC}}$ is given by
\begin{align}
    \hat{H}_{\mathrm{SOC}} = 2D_y\sum_i \left( \hat{S}^z_{l i}  \hat{S}^x_{l i+1} + \hat{S}^z_{r i}  \hat{S}^x_{r i+1}\right).
\end{align}
We can now define a flux operator on each plaquette as
\begin{align}
    \hat{W}_i = 16\hat{S}^x_{li}\hat{S}^x_{ri}\hat{S}^z_{l i+1}\hat{S}^z_{l i+1}.
\end{align}
with eigenvalues $\pm1$. Because $\hat{W}_i$ commutes with $\hat{H}$, the total Hilbert space splits into exponentially many pieces labeled by the eigenvalues of $\hat{W}_i$. This effect is called local Hilbert space fragmentation~\cite{buvca2022out} and is linked to nonergodic behavior of many-body states, i.e., ``slower than usual" thermalization~\cite{moudgalya2022quantum}. For a finite system of size $L$, there are additional conserved charges given by the operators $\hat{W}_{\mathrm{top}}=4\hat{S}^z_{l1}\hat{S}^z_{r 1}$ and $\hat{W}_{\mathrm{bottom}}=4\hat{S}^x_{lL}\hat{S}^x_{r L}$. 

The presence of these conserved charges is a property of the model, and while it certainly deserves further investigation, it is not the focus of this manuscript and we leave it to future research to understand its unusual time evolution.

\bibliographystyle{unsrt}
\bibliography{references}

\begin{thebibliography}{100}

\bibitem{hosten2008observation}
Onur Hosten and Paul Kwiat.
\newblock Observation of the spin hall effect of light via weak measurements.
\newblock {\em Science}, 319(5864):787--790, 2008.

\bibitem{wang2009observation}
Zheng Wang, Yidong Chong, John~D Joannopoulos, and Marin Solja{\v{c}}i{\'c}.
\newblock Observation of unidirectional backscattering-immune topological electromagnetic states.
\newblock {\em Nature}, 461(7265):772--775, 2009.

\bibitem{ben2016photon}
Philippe Ben-Abdallah.
\newblock Photon thermal hall effect.
\newblock {\em Physical Review Letters}, 116(8):084301, 2016.

\bibitem{haldane2008possible}
Frederick Duncan~Michael Haldane and Srinivas Raghu.
\newblock Possible realization of directional optical waveguides in photonic crystals with broken time-reversal symmetry.
\newblock {\em Physical Review Letters}, 100(1):013904, 2008.

\bibitem{raghu2008analogs}
Srinivas Raghu and Frederick Duncan~Michael Haldane.
\newblock Analogs of quantum-hall-effect edge states in photonic crystals.
\newblock {\em Physical Review A}, 78(3):033834, 2008.

\bibitem{onoda2004hall}
Masaru Onoda, Shuichi Murakami, and Naoto Nagaosa.
\newblock Hall effect of light.
\newblock {\em Physical review letters}, 93(8):083901, 2004.

\bibitem{strohm2005phenomenological}
C~Strohm, GLJA Rikken, and P~Wyder.
\newblock Phenomenological evidence for the phonon hall effect.
\newblock {\em Physical Review Letters}, 95(15):155901, 2005.

\bibitem{kagan2008anomalous}
Yu~Kagan and LA~Maksimov.
\newblock Anomalous hall effect for the phonon heat conductivity in paramagnetic dielectrics.
\newblock {\em Physical Review Letters}, 100(14):145902, 2008.

\bibitem{sugii2017thermal}
Kaori Sugii, Masaaki Shimozawa, Daiki Watanabe, Yoshitaka Suzuki, Mario Halim, Motoi Kimata, Yosuke Matsumoto, Satoru Nakatsuji, and Minoru Yamashita.
\newblock Thermal hall effect in a phonon-glass ba 3 cusb 2 o 9.
\newblock {\em Physical Review Letters}, 118(14):145902, 2017.

\bibitem{sheng2006theory}
L~Sheng, DN~Sheng, and CS~Ting.
\newblock Theory of the phonon hall effect in paramagnetic dielectrics.
\newblock {\em Physical Review Letters}, 96(15):155901, 2006.

\bibitem{katsura_theory_of_thermal_hall}
Hosho Katsura, Naoto Nagaosa, and Patrick~A. Lee.
\newblock Theory of the thermal hall effect in quantum magnets.
\newblock {\em Phys. Rev. Lett.}, 104:066403, Feb 2010.

\bibitem{kondo2019three}
Hiroki Kondo, Yutaka Akagi, and Hosho Katsura.
\newblock Three-dimensional topological magnon systems.
\newblock {\em Physical Review B}, 100(14):144401, 2019.

\bibitem{kondo2021dirac}
Hiroki Kondo and Yutaka Akagi.
\newblock Dirac surface states in magnonic analogs of topological crystalline insulators.
\newblock {\em Physical Review Letters}, 127(17):177201, 2021.

\bibitem{matsumoto2011rotational}
Ryo Matsumoto and Shuichi Murakami.
\newblock Rotational motion of magnons and the thermal hall effect.
\newblock {\em Physical Review B}, 84(18):184406, 2011.

\bibitem{fujimoto2009hall}
Satoshi Fujimoto.
\newblock Hall effect of spin waves in frustrated magnets.
\newblock {\em Physical Review Letters}, 103(4):047203, 2009.

\bibitem{shindou2013chiral}
Ryuichi Shindou, Jun-ichiro Ohe, Ryo Matsumoto, Shuichi Murakami, and Eiji Saitoh.
\newblock Chiral spin-wave edge modes in dipolar magnetic thin films.
\newblock {\em Physical Review B}, 87(17):174402, 2013.

\bibitem{shindou2013topological}
Ryuichi Shindou, Ryo Matsumoto, Shuichi Murakami, and Jun-ichiro Ohe.
\newblock Topological chiral magnonic edge mode in a magnonic crystal.
\newblock {\em Physical Review B}, 87(17):174427, 2013.

\bibitem{kim2016realization}
Se~Kwon Kim, H{\'e}ctor Ochoa, Ricardo Zarzuela, and Yaroslav Tserkovnyak.
\newblock Realization of the haldane-kane-mele model in a system of localized spins.
\newblock {\em Physical Review Letters}, 117(22):227201, 2016.

\bibitem{joshi2018KH}
Darshan~G. Joshi.
\newblock Topological excitations in the ferromagnetic kitaev-heisenberg model.
\newblock {\em Phys. Rev. B}, 98:060405, Aug 2018.

\bibitem{romhanyi2015hall}
Judit Romh{\'a}nyi, Karlo Penc, and Ramachandran Ganesh.
\newblock Hall effect of triplons in a dimerized quantum magnet.
\newblock {\em Nature communications}, 6(1):6805, 2015.

\bibitem{joshi2017topological}
Darshan~G Joshi and Andreas~P Schnyder.
\newblock Topological quantum paramagnet in a quantum spin ladder.
\newblock {\em Physical Review B}, 96(22):220405, 2017.

\bibitem{joshi2019z2}
Darshan~G. Joshi and Andreas~P. Schnyder.
\newblock ${\mathbb{z}}_{2}$ topological quantum paramagnet on a honeycomb bilayer.
\newblock {\em Phys. Rev. B}, 100:020407, Jul 2019.

\bibitem{chumak2015magnon}
Andrii~V Chumak, Vitaliy~I Vasyuchka, Alexander~A Serga, and Burkard Hillebrands.
\newblock Magnon spintronics.
\newblock {\em Nature physics}, 11(6):453--461, 2015.

\bibitem{li2021topological_applications}
Z-X Li, Yunshan Cao, and Peng Yan.
\newblock Topological insulators and semimetals in classical magnetic systems.
\newblock {\em Physics Reports}, 915:1--64, 2021.

\bibitem{matsumoto2011theoretical}
Ryo Matsumoto and Shuichi Murakami.
\newblock Theoretical prediction of a rotating magnon wave packet in ferromagnets.
\newblock {\em Physical Review Letters}, 106(19):197202, 2011.

\bibitem{wang2018topological}
XS~Wang, HW~Zhang, and XR~Wang.
\newblock Topological magnonics: A paradigm for spin-wave manipulation and device design.
\newblock {\em Physical Review Applied}, 9(2):024029, 2018.

\bibitem{costa2020topological}
Ant{\'o}nio~T Costa, Daniel Louren{\c{c}}o~R Santos, Nuno~MR Peres, and Joaqu{\'\i}n Fern{\'a}ndez-Rossier.
\newblock Topological magnons in cri3 monolayers: an itinerant fermion description.
\newblock {\em 2D Materials}, 7(4):045031, 2020.

\bibitem{li2021topological}
Shuyi Li and Andriy~H Nevidomskyy.
\newblock Topological weyl magnons and thermal hall effect in layered honeycomb ferromagnets.
\newblock {\em Physical Review B}, 104(10):104419, 2021.

\bibitem{hwang2020topological}
Kyusung Hwang, Nandini Trivedi, and Mohit Randeria.
\newblock Topological magnons with nodal-line and triple-point degeneracies: implications for thermal hall effect in pyrochlore iridates.
\newblock {\em Physical Review Letters}, 125(4):047203, 2020.

\bibitem{laurell2018magnon}
Pontus Laurell and Gregory~A Fiete.
\newblock Magnon thermal hall effect in kagome antiferromagnets with dzyaloshinskii-moriya interactions.
\newblock {\em Physical Review B}, 98(9):094419, 2018.

\bibitem{mook2014magnon}
Alexander Mook, J{\"u}rgen Henk, and Ingrid Mertig.
\newblock Magnon hall effect and topology in kagome lattices: A theoretical investigation.
\newblock {\em Physical Review B}, 89(13):134409, 2014.

\bibitem{owerre2016topological}
SA~Owerre.
\newblock Topological honeycomb magnon hall effect: A calculation of thermal hall conductivity of magnetic spin excitations.
\newblock {\em Journal of Applied Physics}, 120(4), 2016.

\bibitem{han2017spin}
Jung~Hoon Han and Hyunyong Lee.
\newblock Spin chirality and hall-like transport phenomena of spin excitations.
\newblock {\em Journal of the Physical Society of Japan}, 86(1):011007, 2017.

\bibitem{li2016weyl}
Fei-Ye Li, Yao-Dong Li, Yong~Baek Kim, Leon Balents, Yue Yu, and Gang Chen.
\newblock Weyl magnons in breathing pyrochlore antiferromagnets.
\newblock {\em Nature communications}, 7(1):12691, 2016.

\bibitem{zhu2021topological}
Fengfeng Zhu, Lichuan Zhang, Xiao Wang, Flaviano~Jos{\'e} Dos~Santos, Junda Song, Thomas Mueller, Karin Schmalzl, Wolfgang~F Schmidt, Alexandre Ivanov, Jitae~T Park, et~al.
\newblock Topological magnon insulators in two-dimensional van der waals ferromagnets crsite3 and crgete3: Toward intrinsic gap-tunability.
\newblock {\em Science advances}, 7(37):eabi7532, 2021.

\bibitem{zhang2021anomalous}
Heda Zhang, Chunqiang Xu, Caitlin Carnahan, Milos Sretenovic, Nishchay Suri, Di~Xiao, and Xianglin Ke.
\newblock Anomalous thermal hall effect in an insulating van der waals magnet.
\newblock {\em Physical Review Letters}, 127(24):247202, 2021.

\bibitem{mcclarty2017topological}
Paul~A McClarty, F~Kr{\"u}ger, Tatiana Guidi, SF~Parker, Keith Refson, AW~Parker, Dharmalingam Prabhakaran, and Radu Coldea.
\newblock Topological triplon modes and bound states in a shastry--sutherland magnet.
\newblock {\em Nature Physics}, 13(8):736--741, 2017.

\bibitem{czajka2023planar}
Peter Czajka, Tong Gao, Max Hirschberger, Paula Lampen-Kelley, Arnab Banerjee, Nicholas Quirk, David~G Mandrus, Stephen~E Nagler, and N~Phuan Ong.
\newblock Planar thermal hall effect of topological bosons in the kitaev magnet $\alpha$-rucl3.
\newblock {\em Nature Materials}, 22(1):36--41, 2023.

\bibitem{hirschberger2015thermal}
Max Hirschberger, Robin Chisnell, Young~S Lee, and Nai~Phuan Ong.
\newblock Thermal hall effect of spin excitations in a kagome magnet.
\newblock {\em Physical Review Letters}, 115(10):106603, 2015.

\bibitem{onose2010observation}
Y~Onose, T~Ideue, H~Katsura, Y~Shiomi, N~Nagaosa, and Y~Tokura.
\newblock Observation of the magnon hall effect.
\newblock {\em Science}, 329(5989):297--299, 2010.

\bibitem{suetsugu2022intrinsic}
S~Suetsugu, T~Yokoi, K~Totsuka, T~Ono, I~Tanaka, S~Kasahara, Y~Kasahara, Z~Chengchao, H~Kageyama, and Y~Matsuda.
\newblock Intrinsic suppression of the topological thermal hall effect in an exactly solvable quantum magnet.
\newblock {\em Physical Review B}, 105(2):024415, 2022.

\bibitem{cairns2020thermal}
Luke~Pritchard Cairns, J-Ph Reid, Robin Perry, Dharmalingam Prabhakaran, and Andrew Huxley.
\newblock Thermal hall effect of topological triplons in srcu2 (bo3) 2.
\newblock In {\em Proceedings of the International Conference on Strongly Correlated Electron Systems (SCES2019)}, page 011089, 2020.

\bibitem{chernyshev_review}
ME~Zhitomirsky and AL~Chernyshev.
\newblock Colloquium: Spontaneous magnon decays.
\newblock {\em Reviews of Modern Physics}, 85(1):219, 2013.

\bibitem{chernyshev2016damped}
AL~Chernyshev and PA~Maksimov.
\newblock Damped topological magnons in the kagome-lattice ferromagnets.
\newblock {\em Physical Review Letters}, 117(18):187203, 2016.

\bibitem{habel2023breakdown}
Jonas Habel, Alexander Mook, Josef Willsher, and Johannes Knolle.
\newblock Breakdown of chiral edge modes in topological magnon insulators.
\newblock {\em arXiv preprint arXiv:2308.03168}, 2023.

\bibitem{thomasen2021fragility}
Andreas Thomasen, Karlo Penc, Nic Shannon, and Judit Romh{\'a}nyi.
\newblock Fragility of z 2 topological invariant characterizing triplet excitations in a bilayer kagome magnet.
\newblock {\em Physical Review B}, 104(10):104412, 2021.

\bibitem{dirac_magnon_honeycomb_ferromagnets_interaction}
Sergey~S. Pershoguba, Saikat Banerjee, J.~C. Lashley, Jihwey Park, Hans \AA{}gren, Gabriel Aeppli, and Alexander~V. Balatsky.
\newblock Dirac magnons in honeycomb ferromagnets.
\newblock {\em Phys. Rev. X}, 8:011010, Jan 2018.

\bibitem{winter2017breakdown}
Stephen~M Winter, Kira Riedl, Pavel~A Maksimov, Alexander~L Chernyshev, Andreas Honecker, and Roser Valent{\'\i}.
\newblock Breakdown of magnons in a strongly spin-orbital coupled magnet.
\newblock {\em Nature communications}, 8(1):1152, 2017.

\bibitem{sengupta}
Hao Sun, Dhiman Bhowmick, Bo~Yang, and Pinaki Sengupta.
\newblock Interacting topological dirac magnons.
\newblock {\em Phys. Rev. B}, 107:134426, Apr 2023.

\bibitem{mcclarty_topological_magnons}
P.~A. McClarty, X.-Y. Dong, M.~Gohlke, J.~G. Rau, F.~Pollmann, R.~Moessner, and K.~Penc.
\newblock Topological magnons in kitaev magnets at high fields.
\newblock {\em Phys. Rev. B}, 98:060404, Aug 2018.

\bibitem{schmidiger2012spectral}
D~Schmidiger, Pierre Bouillot, S~M{\"u}hlbauer, S~Gvasaliya, Corinna Kollath, Thierry Giamarchi, and A~Zheludev.
\newblock Spectral and thermodynamic properties of a strong-leg quantum spin ladder.
\newblock {\em Physical review letters}, 108(16):167201, 2012.

\bibitem{dagotto1996surprises}
Elbio Dagotto and TM~Rice.
\newblock Surprises on the way from one-to two-dimensional quantum magnets: The ladder materials.
\newblock {\em Science}, 271(5249):618--623, 1996.

\bibitem{sachdev1990bond}
Subir Sachdev and RN~Bhatt.
\newblock Bond-operator representation of quantum spins: Mean-field theory of frustrated quantum heisenberg antiferromagnets.
\newblock {\em Physical Review B}, 41(13):9323, 1990.

\bibitem{larged1}
Darshan~G. Joshi, Kris Coester, Kai~P. Schmidt, and Matthias Vojta.
\newblock Nonlinear bond-operator theory and $1/d$ expansion for coupled-dimer magnets. i. paramagnetic phase.
\newblock {\em Phys. Rev. B}, 91:094404, Mar 2015.

\bibitem{larged2}
Darshan~G. Joshi and Matthias Vojta.
\newblock Nonlinear bond-operator theory and $1/d$ expansion for coupled-dimer magnets. ii. antiferromagnetic phase and quantum phase transition.
\newblock {\em Phys. Rev. B}, 91:094405, Mar 2015.

\bibitem{haldane1988model}
F~Duncan~M Haldane.
\newblock Model for a quantum hall effect without landau levels: Condensed-matter realization of the" parity anomaly".
\newblock {\em Physical Review Letters}, 61(18):2015, 1988.

\bibitem{kane-mele}
C.~L. Kane and E.~J. Mele.
\newblock Quantum spin hall effect in graphene.
\newblock {\em Phys. Rev. Lett.}, 95:226801, Nov 2005.

\bibitem{fu-kane}
Liang Fu, C.~L. Kane, and E.~J. Mele.
\newblock Topological insulators in three dimensions.
\newblock {\em Phys. Rev. Lett.}, 98:106803, Mar 2007.

\bibitem{bhz}
B.~Andrei Bernevig, Taylor~L. Hughes, and Shou-Cheng Zhang.
\newblock Quantum spin hall effect and topological phase transition in hgte quantum wells.
\newblock {\em Science}, 314(5806):1757--1761, 2006.

\bibitem{kondo2019z}
Hiroki Kondo, Yutaka Akagi, and Hosho Katsura.
\newblock Z 2 topological invariant for magnon spin hall systems.
\newblock {\em Physical Review B}, 99(4):041110, 2019.

\bibitem{mentre2006structural}
Olivier Mentr{\'e}, El~Mostafa Ketatni, Marie Colmont, Marielle Huve, Francis Abraham, and Vaclav Petricek.
\newblock Structural features of the modulated bicu2 (p1-x v x) o6 solid solution; 4-d treatment of x= 0.87 compound and magnetic spin-gap to gapless transition in new cu2+ two-leg ladder systems.
\newblock {\em Journal of the American Chemical Society}, 128(33):10857--10867, 2006.

\bibitem{mila1996exchange}
Fr{\'e}d{\'e}ric Mila, Patrice Millet, and Jacques Bonvoisin.
\newblock Exchange integrals of vanadates as revealed by magnetic-susceptibility measurements of nav 2 o 5.
\newblock {\em Physical Review B}, 54(17):11925, 1996.

\bibitem{isobe1996magnetic}
Masahiko Isobe and Yutaka Ueda.
\newblock Magnetic susceptibility of quasi-one-dimensional compound $\alpha$'-n a v 2 o 5--possible spin-peierls compound with high critical temperature of 34 k--.
\newblock {\em Journal of the Physical Society of Japan}, 65(5):1178--1181, 1996.

\bibitem{hiroi1995absence}
Z~Hiroi and M~Takano.
\newblock Absence of superconductivity in the doped antiferromagnetic spin-ladder compound (la, sr) cuo2. 5.
\newblock {\em Nature}, 377(6544):41--43, 1995.

\bibitem{kohama2012anisotropic}
Yoshimitsu Kohama, Shuang Wang, Atsuko Uchida, Krunoslav Prsa, Sergei Zvyagin, Yuri Skourski, Ross~D McDonald, Luis Balicas, Henrik~M Ronnow, Christian R{\"u}egg, et~al.
\newblock Anisotropic cascade of field-induced phase transitions in the frustrated spin-ladder system bicu 2 po 6.
\newblock {\em Physical Review Letters}, 109(16):167204, 2012.

\bibitem{kiryukhin2001magnetic}
V~Kiryukhin, YJ~Kim, KJ~Thomas, FC~Chou, RW~Erwin, Q~Huang, MA~Kastner, and RJ~Birgeneau.
\newblock Magnetic properties of the s= 1 2 quasi-one-dimensional antiferromagnet cacu 2 o 3.
\newblock {\em Physical Review B}, 63(14):144418, 2001.

\bibitem{azuma1994observation}
Masaki Azuma, Z~Hiroi, M~Takano, K~Ishida, and Y~Kitaoka.
\newblock Observation of a spin gap in sr cu 2 o 3 comprising spin-$1/2$ quasi-1d two-leg ladders.
\newblock {\em Physical Review Letters}, 73(25):3463, 1994.

\bibitem{wang2016tunable}
Ya-ping Wang, Chang-wen Zhang, Wei-xiao Ji, Run-wu Zhang, Ping Li, Pei-ji Wang, Miao-juan Ren, Xin-lian Chen, and Min Yuan.
\newblock Tunable quantum spin hall effect via strain in two-dimensional arsenene monolayer.
\newblock {\em Journal of Physics D: Applied Physics}, 49(5):055305, 2016.

\bibitem{cheung2016spin}
Chi-Ho Cheung, Huei-Ru Fuh, Ming-Chien Hsu, Yeu-Chung Lin, and Ching-Ray Chang.
\newblock Spin orbit coupling gap and indirect gap in strain-tuned topological insulator-antimonene.
\newblock {\em Nanoscale Research Letters}, 11:1--8, 2016.

\bibitem{winterfeld2013strain}
Lars Winterfeld, Luis~A Agapito, Jin Li, Nicholas Kioussis, Peter Blaha, and Yong~P Chen.
\newblock Strain-induced topological insulator phase transition in hgse.
\newblock {\em Physical Review B}, 87(7):075143, 2013.

\bibitem{liu2021semimetallic}
Tsai-Jung Liu, Maximilian~A Springer, Niclas Heinsdorf, Agnieszka Kuc, Roser Valent{\'\i}, and Thomas Heine.
\newblock Semimetallic square-octagon two-dimensional polymer with high mobility.
\newblock {\em Physical Review B}, 104(20):205419, 2021.

\bibitem{esaki2023electric}
Nanse Esaki, Yutaka Akagi, and Hosho Katsura.
\newblock Electric field induced thermal hall effect of triplons in the quantum dimer magnets $ x $ cucl $ \_ $\{$3$\}$ $($ x= $ tl, k).
\newblock {\em arXiv preprint arXiv:2309.12812}, 2023.

\bibitem{Sturm+1993+233+242}
K.~Sturm.
\newblock Dynamic structure factor: An introduction.
\newblock {\em Zeitschrift für Naturforschung A}, 48(1-2):233--242, 1993.

\bibitem{van1954correlations}
L{\'e}on Van~Hove.
\newblock Correlations in space and time and born approximation scattering in systems of interacting particles.
\newblock {\em Physical Review}, 95(1):249, 1954.

\bibitem{joshi2018detecting}
Darshan~G Joshi, Andreas~P Schnyder, and So~Takei.
\newblock Detecting end states of topological quantum paramagnets via spin hall noise spectroscopy.
\newblock {\em Physical Review B}, 98(6):064401, 2018.

\bibitem{white1992density}
Steven~R White.
\newblock Density matrix formulation for quantum renormalization groups.
\newblock {\em Physical Review Letters}, 69(19):2863, 1992.

\bibitem{vidal2004efficient}
Guifr{\'e} Vidal.
\newblock Efficient simulation of one-dimensional quantum many-body systems.
\newblock {\em Physical Review Letters}, 93(4):040502, 2004.

\bibitem{daley2004time}
Andrew~John Daley, Corinna Kollath, Ulrich Schollw{\"o}ck, and Guifr{\'e} Vidal.
\newblock Time-dependent density-matrix renormalization-group using adaptive effective hilbert spaces.
\newblock {\em Journal of Statistical Mechanics: Theory and Experiment}, 2004(04):P04005, 2004.

\bibitem{white2004real}
Steven~R White and Adrian~E Feiguin.
\newblock Real-time evolution using the density matrix renormalization group.
\newblock {\em Physical review letters}, 93(7):076401, 2004.

\bibitem{haegeman2011time}
Jutho Haegeman, J~Ignacio Cirac, Tobias~J Osborne, Iztok Pi{\v{z}}orn, Henri Verschelde, and Frank Verstraete.
\newblock Time-dependent variational principle for quantum lattices.
\newblock {\em Physical Review Letters}, 107(7):070601, 2011.

\bibitem{haegeman2016unifying}
Jutho Haegeman, Christian Lubich, Ivan Oseledets, Bart Vandereycken, and Frank Verstraete.
\newblock Unifying time evolution and optimization with matrix product states.
\newblock {\em Physical Review B}, 94(16):165116, 2016.

\bibitem{zaletel2015time}
Michael~P Zaletel, Roger~SK Mong, Christoph Karrasch, Joel~E Moore, and Frank Pollmann.
\newblock Time-evolving a matrix product state with long-ranged interactions.
\newblock {\em Physical Review B}, 91(16):165112, 2015.

\bibitem{footnote_oddpart}
{\color{blue}Odd refers to a sign change under exchange of the two atoms in the unit cell. In a scattering experiment on a stack of ladders, this part of the response can be discriminated by its momentum-dependence along the stacking direction, and for the model considered here it exactly corresponds to the triplon modes $t_x$ and $t_z$ (see Eq. (11), (13)).}

\bibitem{gohlke2018dynamical}
Matthias Gohlke, Roderich Moessner, and Frank Pollmann.
\newblock Dynamical and topological properties of the kitaev model in a [111] magnetic field.
\newblock {\em Physical Review B}, 98(1):014418, 2018.

\bibitem{sherman2023spectral}
Nicholas~E Sherman, Maxime Dupont, and Joel~E Moore.
\newblock Spectral function of the j 1- j 2 heisenberg model on the triangular lattice.
\newblock {\em Physical Review B}, 107(16):165146, 2023.

\bibitem{bouillot2011statics}
Pierre Bouillot, Corinna Kollath, Andreas~M L{\"a}uchli, Mikhail Zvonarev, Benedikt Thielemann, Christian R{\"u}egg, Edmond Orignac, Roberta Citro, Martin Klanj{\v{s}}ek, Claude Berthier, et~al.
\newblock Statics and dynamics of weakly coupled antiferromagnetic spin-1 2 ladders in a magnetic field.
\newblock {\em Physical Review B}, 83(5):054407, 2011.

\bibitem{laurell2022magnetic}
Pontus Laurell, Allen Scheie, D~Alan Tennant, Satoshi Okamoto, Gonzalo Alvarez, and Elbio Dagotto.
\newblock Magnetic excitations, nonclassicality, and quantum wake spin dynamics in the hubbard chain.
\newblock {\em Physical Review B}, 106(8):085110, 2022.

\bibitem{supp}
supplemental.

\bibitem{iraola2021towards}
Mikel Iraola, Niclas Heinsdorf, Apoorv Tiwari, Dominik Lessnich, Thomas Mertz, Francesco Ferrari, Mark~H Fischer, Stephen~M Winter, Frank Pollmann, Titus Neupert, et~al.
\newblock Towards a topological quantum chemistry description of correlated systems: The case of the hubbard diamond chain.
\newblock {\em Physical Review B}, 104(19):195125, 2021.

\bibitem{Verresen2019}
Ruben Verresen, Roderich Moessner, and Frank Pollmann.
\newblock Avoided quasiparticle decay from strong quantum interactions.
\newblock {\em Nature Physics}, 15(8):750--753, Aug 2019.

\bibitem{alex_interaction_stabilized}
Alexander Mook, Kirill Plekhanov, Jelena Klinovaja, and Daniel Loss.
\newblock Interaction-stabilized topological magnon insulator in ferromagnets.
\newblock {\em Phys. Rev. X}, 11:021061, Jun 2021.

\bibitem{tremblay2008refresher}
Andr{\'e}-Marie Tremblay.
\newblock A refresher in many-body theory.
\newblock {\em CIFAR-PITP Interna-tional Summer School on Numerical Methods for CorrelatedSystems in Condensed Matter, Sherbrooke}, 2008.

\bibitem{fateflatiron}
A.~Hunter, S.~Beck, E.~Cappelli, F.~Margot, M.~Straub, Y.~Alexanian, G.~Gatti, M.~D. Watson, T.~K. Kim, C.~Cacho, N.~C. Plumb, M.~Shi, M.~Radovi\ifmmode~\acute{c}\else \'{c}\fi{}, D.~A. Sokolov, A.~P. Mackenzie, M.~Zingl, J.~Mravlje, A.~Georges, F.~Baumberger, and A.~Tamai.
\newblock Fate of quasiparticles at high temperature in the correlated metal ${\mathrm{sr}}_{2}{\mathrm{ruo}}_{4}$.
\newblock {\em Phys. Rev. Lett.}, 131:236502, Dec 2023.

\bibitem{samajdar2019enhanced}
Rhine Samajdar, Mathias~S Scheurer, Shubhayu Chatterjee, Haoyu Guo, Cenke Xu, and Subir Sachdev.
\newblock Enhanced thermal hall effect in the square-lattice n{\'e}el state.
\newblock {\em Nature Physics}, 15(12):1290--1294, 2019.

\bibitem{kawano2019thermal}
Masataka Kawano and Chisa Hotta.
\newblock Thermal hall effect and topological edge states in a square-lattice antiferromagnet.
\newblock {\em Physical Review B}, 99(5):054422, 2019.

\bibitem{cao2015magnon}
Xiaodong Cao, Kai Chen, and Dahai He.
\newblock Magnon hall effect on the lieb lattice.
\newblock {\em Journal of Physics: Condensed Matter}, 27(16):166003, 2015.

\bibitem{shastry1981exact}
B~Sriram Shastry and Bill Sutherland.
\newblock Exact ground state of a quantum mechanical antiferromagnet.
\newblock {\em Physica B+ C}, 108(1-3):1069--1070, 1981.

\bibitem{strong_triplon_dampening}
M.~E. Zayed, Ch. R\"uegg, Th. Str\"assle, U.~Stuhr, B.~Roessli, M.~Ay, J.~Mesot, P.~Link, E.~Pomjakushina, M.~Stingaciu, K.~Conder, and H.~M. R\o{}nnow.
\newblock Correlated decay of triplet excitations in the shastry-sutherland compound ${\mathbf{\text{srcu}}}_{2}({\mathrm{bo}}_{3}{)}_{2}$.
\newblock {\em Phys. Rev. Lett.}, 113:067201, Aug 2014.

\bibitem{thermal_evolution}
S.~E. Nikitin, B.~F\aa{}k, K.~W. Kr\"amer, T.~Fennell, B.~Normand, A.~M. L\"auchli, and Ch. R\"uegg.
\newblock Thermal evolution of dirac magnons in the honeycomb ferromagnet ${\mathrm{crbr}}_{3}$.
\newblock {\em Phys. Rev. Lett.}, 129:127201, Sep 2022.

\bibitem{temperature_induced_phase_transition}
Yu-Shan Lu, Jian-Lin Li, and Chien-Te Wu.
\newblock Topological phase transitions of dirac magnons in honeycomb ferromagnets.
\newblock {\em Phys. Rev. Lett.}, 127:217202, Nov 2021.

\bibitem{guo2022quasi}
Chunyu Guo, Lunhui Hu, Carsten Putzke, Jonas Diaz, Xiangwei Huang, Kaustuv Manna, Feng-Ren Fan, Chandra Shekhar, Yan Sun, Claudia Felser, et~al.
\newblock Quasi-symmetry-protected topology in a semi-metal.
\newblock {\em Nature physics}, 18(7):813--818, 2022.

\bibitem{shang2023irf}
Chao Shang, Owen Ganter, Niclas Heinsdorf, and Stephen~M Winter.
\newblock Irf 4: From tetrahedral compass model to topological semimetal.
\newblock {\em Physical Review B}, 107(12):125111, 2023.

\bibitem{gurarie2011single}
Victor Gurarie.
\newblock Single-particle green’s functions and interacting topological insulators.
\newblock {\em Physical Review B}, 83(8):085426, 2011.

\bibitem{wang2013topological}
Zhong Wang and Binghai Yan.
\newblock Topological hamiltonian as an exact tool for topological invariants.
\newblock {\em Journal of Physics: Condensed Matter}, 25(15):155601, 2013.

\bibitem{wang2012simplified}
Zhong Wang and Shou-Cheng Zhang.
\newblock Simplified topological invariants for interacting insulators.
\newblock {\em Physical Review X}, 2(3):031008, 2012.

\bibitem{kapustin2020higher}
Anton Kapustin and Lev Spodyneiko.
\newblock Higher-dimensional generalizations of berry curvature.
\newblock {\em Physical Review B}, 101(23):235130, 2020.

\bibitem{shiozaki2023higher}
Ken Shiozaki, Niclas Heinsdorf, and Shuhei Ohyama.
\newblock Higher berry curvature from matrix product states.
\newblock {\em arXiv preprint arXiv:2305.08109}, 2023.

\bibitem{hauke2016measuring}
Philipp Hauke, Markus Heyl, Luca Tagliacozzo, and Peter Zoller.
\newblock Measuring multipartite entanglement through dynamic susceptibilities.
\newblock {\em Nature Physics}, 12(8):778--782, 2016.

\bibitem{hauschild2018efficient}
Johannes Hauschild and Frank Pollmann.
\newblock Efficient numerical simulations with tensor networks: Tensor network python (tenpy).
\newblock {\em SciPost Physics Lecture Notes}, page 005, 2018.

\bibitem{harris2020array}
Charles~R. Harris, K.~Jarrod Millman, St{\'{e}}fan~J. van~der Walt, Ralf Gommers, Pauli Virtanen, David Cournapeau, Eric Wieser, Julian Taylor, Sebastian Berg, Nathaniel~J. Smith, Robert Kern, Matti Picus, Stephan Hoyer, Marten~H. van Kerkwijk, Matthew Brett, Allan Haldane, Jaime~Fern{\'{a}}ndez del R{\'{i}}o, Mark Wiebe, Pearu Peterson, Pierre G{\'{e}}rard-Marchant, Kevin Sheppard, Tyler Reddy, Warren Weckesser, Hameer Abbasi, Christoph Gohlke, and Travis~E. Oliphant.
\newblock Array programming with {NumPy}.
\newblock {\em Nature}, 585(7825):357--362, September 2020.

\bibitem{bao2018discovery}
Song Bao, Jinghui Wang, Wei Wang, Zhengwei Cai, Shichao Li, Zhen Ma, Di~Wang, Kejing Ran, Zhao-Yang Dong, DL~Abernathy, et~al.
\newblock Discovery of coexisting dirac and triply degenerate magnons in a three-dimensional antiferromagnet.
\newblock {\em Nature communications}, 9(1):2591, 2018.

\bibitem{elliot2021order}
M~Elliot, Paul~A McClarty, D~Prabhakaran, RD~Johnson, HC~Walker, P~Manuel, and R~Coldea.
\newblock Order-by-disorder from bond-dependent exchange and intensity signature of nodal quasiparticles in a honeycomb cobaltate.
\newblock {\em Nature Communications}, 12(1):3936, 2021.

\bibitem{yao2018topological}
Weiliang Yao, Chenyuan Li, Lichen Wang, Shangjie Xue, Yang Dan, Kazuki Iida, Kazuya Kamazawa, Kangkang Li, Chen Fang, and Yuan Li.
\newblock Topological spin excitations in a three-dimensional antiferromagnet.
\newblock {\em Nature Physics}, 14(10):1011--1015, 2018.

\bibitem{yuan2020dirac}
Bo~Yuan, Ilia Khait, Guo-Jiun Shu, FC~Chou, MB~Stone, JP~Clancy, Arun Paramekanti, and Young-June Kim.
\newblock Dirac magnons in a honeycomb lattice quantum xy magnet cotio 3.
\newblock {\em Physical Review X}, 10(1):011062, 2020.

\bibitem{zhang2020magnonic}
L-C Zhang, YA~Onykiienko, PM~Buhl, YV~Tymoshenko, P~{\v{C}}erm{\'a}k, A~Schneidewind, JR~Stewart, A~Henschel, M~Schmidt, S~Bl{\"u}gel, et~al.
\newblock Magnonic weyl states in cu 2 oseo 3.
\newblock {\em Physical review research}, 2(1):013063, 2020.

\bibitem{gornyi2005interacting}
Igor~V Gornyi, Alexander~D Mirlin, and Dmitry~G Polyakov.
\newblock Interacting electrons in disordered wires: Anderson localization and low-t transport.
\newblock {\em Physical review letters}, 95(20):206603, 2005.

\bibitem{nandkishore2015many}
Rahul Nandkishore and David~A Huse.
\newblock Many-body localization and thermalization in quantum statistical mechanics.
\newblock {\em Annu. Rev. Condens. Matter Phys.}, 6(1):15--38, 2015.

\bibitem{altman2015universal}
Ehud Altman and Ronen Vosk.
\newblock Universal dynamics and renormalization in many-body-localized systems.
\newblock {\em Annu. Rev. Condens. Matter Phys.}, 6(1):383--409, 2015.

\bibitem{alet2018many}
Fabien Alet and Nicolas Laflorencie.
\newblock Many-body localization: An introduction and selected topics.
\newblock {\em Comptes Rendus Physique}, 19(6):498--525, 2018.

\bibitem{abanin2019colloquium}
Dmitry~A Abanin, Ehud Altman, Immanuel Bloch, and Maksym Serbyn.
\newblock Colloquium: Many-body localization, thermalization, and entanglement.
\newblock {\em Reviews of Modern Physics}, 91(2):021001, 2019.

\bibitem{jermyn2014stability}
Adam~S Jermyn, Roger~SK Mong, Jason Alicea, and Paul Fendley.
\newblock Stability of zero modes in parafermion chains.
\newblock {\em Physical Review B}, 90(16):165106, 2014.

\bibitem{kemp2017long}
Jack Kemp, Norman~Y Yao, Christopher~R Laumann, and Paul Fendley.
\newblock Long coherence times for edge spins.
\newblock {\em Journal of Statistical Mechanics: Theory and Experiment}, 2017(6):063105, 2017.

\bibitem{else2017prethermal}
Dominic~V Else, Paul Fendley, Jack Kemp, and Chetan Nayak.
\newblock Prethermal strong zero modes and topological qubits.
\newblock {\em Physical Review X}, 7(4):041062, 2017.

\bibitem{serbyn2021quantum}
Maksym Serbyn, Dmitry~A Abanin, and Zlatko Papi{\'c}.
\newblock Quantum many-body scars and weak breaking of ergodicity.
\newblock {\em Nature Physics}, 17(6):675--685, 2021.

\bibitem{chandran2023quantum}
Anushya Chandran, Thomas Iadecola, Vedika Khemani, and Roderich Moessner.
\newblock Quantum many-body scars: A quasiparticle perspective.
\newblock {\em Annual Review of Condensed Matter Physics}, 14:443--469, 2023.

\bibitem{moudgalya2022quantum}
Sanjay Moudgalya, B~Andrei Bernevig, and Nicolas Regnault.
\newblock Quantum many-body scars and hilbert space fragmentation: a review of exact results.
\newblock {\em Reports on Progress in Physics}, 85(8):086501, 2022.

\bibitem{parker2019topologically}
Daniel~E Parker, Romain Vasseur, and Thomas Scaffidi.
\newblock Topologically protected long edge coherence times in symmetry-broken phases.
\newblock {\em Physical Review Letters}, 122(24):240605, 2019.

\bibitem{amelio2020theory}
Ivan Amelio and Iacopo Carusotto.
\newblock Theory of the coherence of topological lasers.
\newblock {\em Physical Review X}, 10(4):041060, 2020.

\bibitem{zapletal2020long}
Petr Zapletal, Bogdan Galilo, and Andreas Nunnenkamp.
\newblock Long-lived elementary excitations and light coherence in topological lasers.
\newblock {\em Optica}, 7(9):1045--1055, 2020.

\bibitem{bahri2015localization}
Yasaman Bahri, Ronen Vosk, Ehud Altman, and Ashvin Vishwanath.
\newblock Localization and topology protected quantum coherence at the edge of hot matter.
\newblock {\em Nature communications}, 6(1):7341, 2015.

\bibitem{chaunsali2021stability}
Rajesh Chaunsali, Haitao Xu, Jinkyu Yang, Panayotis~G Kevrekidis, and Georgios Theocharis.
\newblock Stability of topological edge states under strong nonlinear effects.
\newblock {\em Physical Review B}, 103(2):024106, 2021.

\bibitem{smirnova2020nonlinear}
Daria Smirnova, Daniel Leykam, Yidong Chong, and Yuri Kivshar.
\newblock Nonlinear topological photonics.
\newblock {\em Applied Physics Reviews}, 7(2), 2020.

\bibitem{leykam2016edge}
Daniel Leykam and Yi~Dong Chong.
\newblock Edge solitons in nonlinear-photonic topological insulators.
\newblock {\em Physical review letters}, 117(14):143901, 2016.

\bibitem{goldman2012detecting}
Nathan Goldman, J{\'e}r{\^o}me Beugnon, and Fabrice Gerbier.
\newblock Detecting chiral edge states in the hofstadter optical lattice.
\newblock {\em Physical review letters}, 108(25):255303, 2012.

\bibitem{goldman2013direct}
Nathan Goldman, Jean Dalibard, Alexandre Dauphin, Fabrice Gerbier, Maciej Lewenstein, Peter Zoller, and Ian~B Spielman.
\newblock Direct imaging of topological edge states in cold-atom systems.
\newblock {\em Proceedings of the National Academy of Sciences}, 110(17):6736--6741, 2013.

\bibitem{yao2017theory}
Yuan Yao, Masahiro Sato, Tetsuya Nakamura, Nobuo Furukawa, and Masaki Oshikawa.
\newblock Theory of electron spin resonance in one-dimensional topological insulators with spin-orbit couplings: Detection of edge states.
\newblock {\em Physical Review B}, 96(20):205424, 2017.

\bibitem{buvca2022out}
Berislav Bu{\v{c}}a.
\newblock Out-of-time-ordered crystals and fragmentation.
\newblock {\em Physical Review Letters}, 128(10):100601, 2022.

\bibitem{schnyder2008classification}
Andreas~P Schnyder, Shinsei Ryu, Akira Furusaki, and Andreas~WW Ludwig.
\newblock Classification of topological insulators and superconductors in three spatial dimensions.
\newblock {\em Physical Review B}, 78(19):195125, 2008.

\end{thebibliography}

\clearpage
\newpage

\setcounter{figure}{0}
\makeatletter
\renewcommand{\fnum@figure}{\figurename~S\thefigure}
\makeatother

\section*{Supplementary Material\label{sec:supp}}

\section{DMRG and Time-Evolution\label{sec:s_dmrgandtime}}

We use DMRG and the $\hat{W}^\RN{2}$ time-evolution method for the data on the 8- and 12-rung ladders, and time-evolving block decimation (TEBD) for the calculations on the 32- and 64-rung systems, as implemented in the TeNPy package~\cite{hauschild2018efficient}. The sampling rate (or time step $\Delta t$) is chosen such that the fastest frequency of the signal is resolved, which is the ground state frequency $\omega_0$. The resolution of the Fourier transformed output along the frequency axis - but also the severity of finite size effects due to reflections at the boundaries - is then determined by the total evolved time $T=N_{\Delta t}\Delta t$, where $N_{\Delta t}$ is the number of time steps.

To reduce ringing effects the correlation functions are convoluted with a hamming window along the time axis, before being Fourier transformed using Fast Fourier Transform (FFT) as implemented in the numpy package~\cite{harris2020array}. To obtain the Fourier transform in real-space, all matrix elements $C_{ij}$ are collected with respect to their relative distance along the ladder $d=i-j$. 

Because the DSF is a response function, it is required to be real and positive. In principle, this can be used to choose a suitable Fourier transform, the output of which will always meet said requirements. In this work, we choose an agnostic, pre-implemented FFT to be able to judge the severity of finite-size and -time effects by keeping track of the deviations from these properties as well as the violation of sum rules. 

\begin{figure}
    \centering
    \includegraphics[width=1\linewidth]{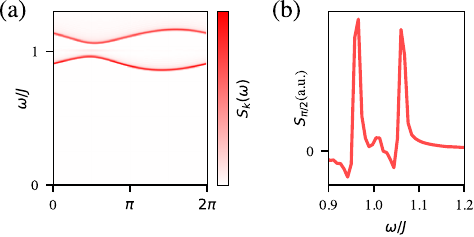}
    \caption{(a) DSF for $K/J=0.01$, $h_y/J=0.05$ and $D_y=\Gamma_y=0.1$ on a 64-rung ladder (128 spin-half sites). (b) Cut through the DSF shown in (a) at $k=\pi/2$. At $\omega/J=1$ there is a quasi-particle peak corresponding to the localized topological boundary mode.}    \label{fig:L64fig}
\end{figure}

Open boundary conditions break translational invariance. Thus, momentum is not a good quantum number. The number of wavelengths output by a discrete FFT only depends on the number of samples in the input signal. For the system sizes considered in this work, we find excellent agreement between these wavelengths and the momenta of an infinite (non-interacting) chain (see Fig.\ 2(a)).

The algorithm that is used to compute the DSF is summarized as follows:

\begin{enumerate}
    \item Compute ground state $|\psi_0 \rangle$ using (finite) DMRG.
    \item Compute \label{item: apply_first_op} $|\psi^{\gamma'}_j(t=0) \rangle = \hat{\tilde{S}}^{\gamma'}_j|\psi_0 \rangle$.
    \item \label{item: time_step}Time-evolve one time step using $\hat{W}^\RN{2}$ or TEBD:\\ \begin{align} 
        \hat{U}(\Delta t)|\psi^{\gamma'}_j(t=0) \rangle = |\psi^{\gamma'}_j(t + \Delta t )\rangle \nonumber
    \end{align}
    \item Compute $|\psi^{\gamma\gamma'}_{ij}(t + \Delta t )\rangle = \hat{\tilde{S}}^{\gamma}_j |\psi^{\gamma}_i(t + \Delta t )\rangle $ for all $i$.
    \item \label{item: overlap} Compute overlap $C^{\gamma \gamma'}_{ij}(t + \Delta t) = \langle \psi_0 |\psi^{\gamma\gamma'}_{ij}(t + \Delta t )\rangle$
    \item \label{item: repeat_times}Repeat steps \ref{item: time_step} - \ref{item: overlap} $N_{\Delta t} = T / \Delta t$ times. 
    \item Repeat steps \ref{item: apply_first_op} - \ref{item: repeat_times} for all $j$. 
    \item Fourier transform $C^{\gamma \gamma'}_{ij}(t)$ in time and sites. 
\end{enumerate}

For the first step, we use a two-site DMRG algorithm as implemented in tenpy\cite{hauschild2018efficient} with convergence criteria on the ground state energy and the entanglement entropy as $\Delta E < 10^{-8}J$ and $\Delta S < 10^{-6}$. For the 8- and 12-rung ladders we choose a matrix product operator (MPO) representation of the Hamiltonian that consists of 16 and 24 spin-half sites respectively. For the 32- and 64-rung systems we group the sites of one rung. At the cost of having a larger dimension of the local Hilbert space, this yields a MPO representation with only nearest-neighbor couplings. For both representations (and both time-evolution methods), we find a bond dimension of $\chi=128$ to be sufficient. DMRG is a variational method, and to avoid a biased intial state we applied random unitaries to the inital MPS.

The main convergence criterion on the time-evolution is the time-step size $\Delta t$. While the overall error for a fixed time-step size is cumulative (it is applied $N_{\Delta t}$ times), convergence can be checked by fixing $T$ and decreasing $\Delta t$. Importantly, the time-step size also fixes the maximal frequency that is resolved. For our system (especially for strong SOC), a choice of $\Delta t$ that is sufficiently small to converge yields a range of $\omega$ that is approximately a hundred times larger than the window that we are interested in, i.e. is the energy range of the edge mode and the upper bulk triplon mode. Therefore, to get a good resolution -- in particular for numerical integration of the DSF -- very long evolutions are required. For the TEBD evolution of the 32-rung systems we choose $\Delta t=0.01/J$ and $N_{\Delta t}=30000$. For the $\hat{W}^\RN{2}$ evolution of the 8- and 12-rung ladders, we find $\Delta t=0.02/J$ and $N_{\Delta t}=5000$ sufficient to identify the quasi-particle peak of the topological edge mode in the DSF, but insufficient for the extraction of the area thereunder. 

For $L=64$, we find to require even smaller time-step sizes which, in turn, require larger values of $N_{\Delta t}$ for a fixed frequency resolution. For $\Delta t=0.01/J$ and $N_{\Delta t}=30000$ (same as for $L=32$), we find the in-gap mode in the DSF as shown in Fig.\ \ref{fig:L64fig}, however the negative dip of the left tail of the lower bulk triplon mode is more severe compared to the 32-rung ladders.

To compute the real space profile of the in-gap mode, we first calculate the local density of states of the lower band
\begin{align}
    \rho_i = \sum_k e^{-ir_jk}\int_0^{J+\delta} d\omega S_k(\omega),
\end{align}
where $\delta$ is a small enough value such that the spectral weight of any boundary mode, but not that of the upper triplon band is included in the integral. We then define the ``topological density of states" as the difference of the local densities of states of the nontrivial paramagnet for open and periodic boundary conditions \cite{joshi2017topological}
\begin{align}
    \rho_i^{\text{top}} = \rho_i^{\text{OBC}} - \rho_i^{\text{PBC}}.
\end{align}
Finally, $\rho_i^{\text{top}}$ is integrated over one half of the ladder to obtain $n_t$ \mbox{(see Eq.\ \ref{eq:nt_integral_formula})}. This method of extracting $n_t$ relies on the cancellation of $\rho_i^{\text{OBC}}$ and $\rho_i^{\text{PBC}}$. The dominating bulk triplon peaks have much higher intensity than the in-gap mode, which means that slightly imperfect cancellation of these contributions translates into strong deviations of $n_t$. Moreover, periodic boundary conditions require long-range couplings in the MPO representation of the Hamiltonian. We use the ``folded" ordering as implemented in tenpy\cite{hauschild2018efficient}. We consider the numerical accuracy sufficient to convince ourselves of the localization of the in-gap modes, as well as the fractionalization in triplon number. We found it difficult to accurately determine $n_t$ for large system sizes that require long time evolution, because over time the difference of open versus periodic boundary conditions accumulates which leads to imperfect cancellation of the two densities of states.

\section{Case (i): Strong Field and Linear Spin Wave Theory}

\begin{figure}
    \centering
    \includegraphics{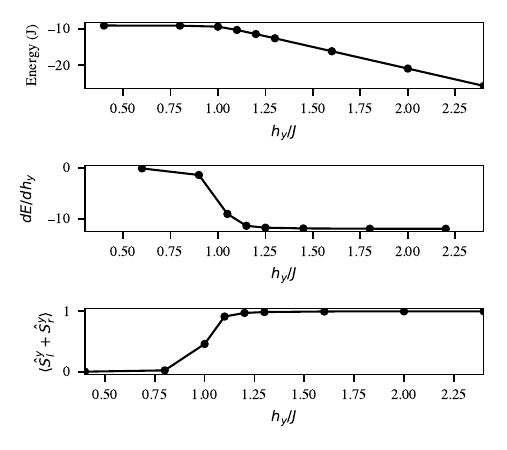}
    \caption{The ground state energy, its derivative and the magnetization along the external field axis as a function of field strength $h_y/J$ for $K/J=0.01$, $D_y/J=\Gamma_y/J=0.1$ and $L=12$. At $h_y/J \sim 1$ there is a phase transition from the paramagnetic to the field-polarized phase.}
    \label{fig:supp_strongfield}
\end{figure}

\begin{figure*}
    \centering
    \includegraphics{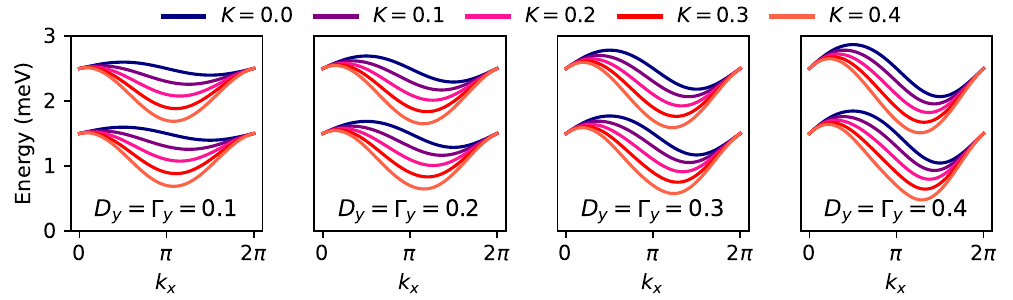}
    \caption{Magnon spectra of the field polarized ladder for $h_y/J = 2.5$ obtained by LSWT for different values of rail coupling $K$ and SOC. The model has two magnetic sites per unit cell and hence two magnon bands. Two bands of the same model are given in the same color. There are no band crossings and the bands are topologically trivial.}
    \label{fig:lswt}
\end{figure*}

For strong external magnetic fields $h_y/J \sim 1$, there is a transition to the field polarized phase as can be seen in Fig.\ S\ref{fig:supp_strongfield}. At the phase transition point, $J$ and $h_y$ are much larger than the other parameters in the model, so the energy decreases approximately linearly with $h_y$. The condensation of the triplons is seen as a jump in the curves derivative, as well as in the order parameter $ \langle\hat{S}^y_{l} + \hat{S}^y_{r}\rangle$, which measures the magnetization along the field axis per dimer. For $K=D_y=\Gamma_y=0$, the transition is more abrupt, because the lower triplon band is completely flat. The ground state can no longer be described as a chain of coupled singlet states. Instead, the ground state is now ferromagnetic, with a classical ground state configuration of all spins pointing along the $y$-direction. This ordered state has magnons as its low-lying excitations. We perform Linear Spin Wave Theory (LSWT), and plot the magnon band structures for a variety of different parameters in Fig.\ S\ref{fig:lswt}. In the parameter regime relevant to our model, no band crossing can be enforced. All sets of the so-obtained magnon bands are adiabatically connected, i.e. they lie in the same topological sector (which is trivial). In contrast to the excitations of the singlet chain, there is no (shifted) chiral symmetry, that could be used to block off-diagonalize the magnon Hamiltonian and define a topological winding number\cite{schnyder2008classification}. We note that the magnon spectrum is gapped, because the Hamiltonian does not have SU(2) symmetry.

\section{Case (ii): Strong Rail Coupling}

\begin{figure}
    \centering
    \includegraphics[width=0.9\linewidth]{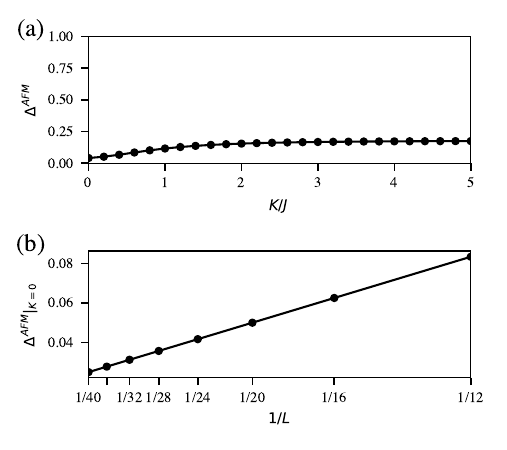}
    \caption{(a) Antiferrromagnetic order parameter for $h_y=0$, $D_y=\Gamma_y=0.1J$ on a ladder with $L=24$ rungs. The order parameter is small, but finite even for $K=0$. (b) Finite size scaling of the order parameter at $K=0$. The order parameter approaches zero in the thermodynamic limit.}
    \label{fig:supp_strongK}
\end{figure}

In the limit of strong rail coupling $K$, the ladder model resembles two weakly coupled anti-ferromagnetic Heisenberg chains. We define the antiferromagnetic order parameter (for the left rail) as 
\begin{align}
    \Delta^{\text{AFM}} = \frac{1}{\left( L/2\right)^2}\sum_{\gamma = x,y,z}\sum_{i,j}^L (-1)^{i+j}\langle \hat{S}^\gamma_{l i}\hat{S}^\gamma_{l j} \rangle/3.
\end{align}
The order parameter for the right rail ($\alpha=r)$ is equivalent. We perform DMRG ground state calculations on a ladder with $L=24$ with $D_y/J=\Gamma_y/J=0.1$ without external magnetic field and a bond dimension cutoff of $\chi=1024$, and sweep through the rail coupling parameter $K$. For finite system size and finite SOC, the order parameter is nonzero. We then do a finite size scaling. For each value of $L$, we compute the ground state and its order parameter for finite $K$, which we then slowly bring to zero, using the MPS from the previous calculation as the initial state until $K=0$. The order parameters of these ``annealed" states as a function of $1/L$ approaches zero in the thermodynamic as shown in Fig. S\ref{fig:supp_strongK}. Hence, the system does not order, as expected for one dimension.

\section{Case (iii): Strong Spin-Orbit Coupling}

\begin{figure*}
    \centering
    \includegraphics{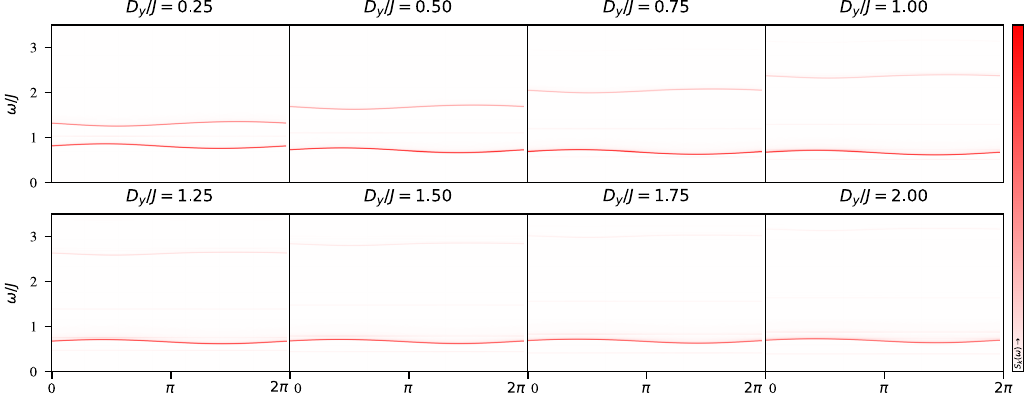}
    \caption{DSF with $K=0.01J$, $h_y = 0.05J$ for different values of $D_y$ and $\Gamma_y$ on a ladder with $L=32$ rungs. The upper triplon branches moves up in energy and loses intensity. The edge mode moves up in energy too, albeit slower than the upper bulk mode.}
    \label{fig:strongSOC_DSF_supp}
\end{figure*}

\begin{figure}
    \centering\includegraphics[width=0.9\linewidth]{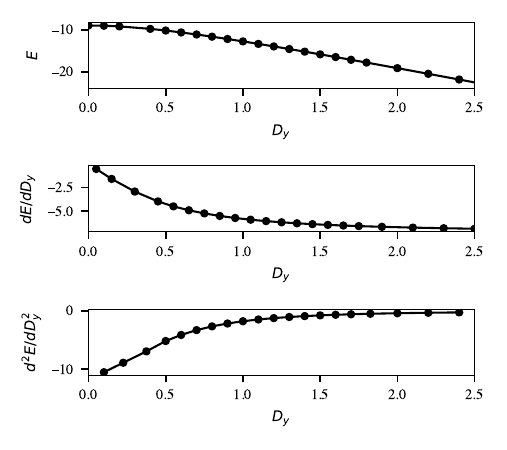}
    \caption{Ground state energy (a) and its first (b) and second (c) derivative with respect to SOC $D_y$ (in units of $J$) with $K=0.01J$, $h_y = 0$ and $\Gamma_y=D_y$ on a ladder with $L=12$ rungs. The curves are smooth and there is no evidence of a phase transition.}
    \label{fig:strongSOC_supp}
\end{figure}

Even though SOC gaps out the two triplon modes, the lower mode does not condense even for large values of $D_y$ and $\Gamma_y$. On a 32-rung ladder (i.e. 64), we compute the DSF and show the evolution of the model's excitations as a function of $D_y$ and $\Gamma_y$ in Fig.\ \ref{fig:strongSOC_DSF_supp}. The DMRG and time-evolution data is obtained using the convergence criteria given in Sec.\ \ref{sec:s_dmrgandtime}. The lower triplon mode initially goes down in energy quickly, until it reaches approximately $0.75J$ where it stagnates even for large values of $D_y$ and $\Gamma_y$. Faint features which are completely flat w.r.t. $k$ appear for $D_y/J>1.5$. As a two-point correlation function, the DSF probes processes involving single-triplons, but in this regime there is no clear notion of single- versus multi-triplon states anymore, and the states that were initially separated for small $D_y/J$ now occupy the same energy range. Potentially, these flat modes have their origin in spectral weight that wants to scatter off these states to lower energies, but instead accumulates at some finite lower bound -- just like the bulk mode that cannot be pushed to arbitrarily low energies by increasing $D_y$ and $\Gamma_y$.

The upper triplon mode keeps going up in energy for larger values of $D_y$ and $\Gamma_y$, while at the same time losing intensity. The boundary mode goes up in energy, but at a slower rate than the upper triplon mode. For a detailed discussion of the intensity and area of its quasi-particle peak (and that of the in-gap mode) we refer to the main text.

To confirm the absence of triplon condensation, we compute the ground state energies of a 12-rung ladder for $K/J=0.01$ and $h_y/J=0$ as a function of $D_y$ and $\Gamma_y$, and look for signs of a phase transition. In Fig. S\ref{fig:strongSOC_supp} the ground state energy and its derivatives are given with no sign of jumps or discontinuities.

\end{document}